%% file: ms.tex
\documentclass[12pt,preprint]{aastex}

\usepackage{natbib}
\usepackage{epsfig,graphicx,color,gensymb}
\usepackage{longtable}

\definecolor{orange}{cmyk}{0,0.4,0.8,0.2}
\definecolor{darkorange}{rgb}{.71,0.21,0.01}
\definecolor{darkgreen}{rgb}{.12,.54,.11}

\usepackage{hyperref}
\hypersetup{pdftex,  
  breaklinks=true,  
  colorlinks=true,
  urlcolor=blue,
  linkcolor=darkorange,
  citecolor=darkgreen,
  }

\newcommand{\x}{{\bf x}}

\newcommand{\Prob}{\mathbf{P}}

 \def\hipp {{\it Hipparcos~}}
\def\nomad {{\it NOMAD~}}

\begin{document}

\shorttitle{ASAS Variable Star Catalog}
\shortauthors{J. W. Richards, et al.}
\title{Construction of a Calibrated Probabilistic Classification Catalog: Application to 50k Variable Sources in the All-Sky Automated Survey}
\author{
Joseph W. Richards\altaffilmark{1,2,*},
Dan L. Starr\altaffilmark{1},
Adam A. Miller\altaffilmark{1},
Joshua S. Bloom\altaffilmark{1},
Nathaniel R. Butler\altaffilmark{3},
Henrik Brink\altaffilmark{1},
Arien Crellin-Quick\altaffilmark{1}
}

\altaffiltext{1}{Astronomy Department, University of California, Berkeley, CA, 94720-3411, USA}
\altaffiltext{2}{Statistics Department, University of California, Berkeley, CA, 94720-7450, USA}
\altaffiltext{3}{School of Earth and Space Exploration, Arizona State University}
\altaffiltext{*}{E-mail: {\tt jwrichar@stat.berkeley.edu}}

\slugcomment{Submitted}

\begin{abstract}
With growing data volumes from synoptic surveys, astronomers necessarily must become more abstracted from the discovery and introspection processes. Given the scarcity of follow-up resources, there is a particularly sharp onus on the frameworks that replace these human roles to provide accurate and well-calibrated probabilistic classification catalogs. Such catalogs inform the subsequent follow-up, allowing consumers to optimize the selection of specific sources for further study and  permitting rigorous treatment of purities and efficiencies for population studies.  Here, we describe a process to produce a probabilistic classification catalog of variability with machine learning from a multi-epoch photometric survey. In addition to producing accurate classifications, we show how to estimate \emph{calibrated} class probabilities, and motivate the importance of probability calibration.  We also introduce a methodology for feature-based anomaly detection, which allows discovery of objects in the survey that do not fit within the predefined class taxonomy.  Finally, we apply these methods to sources observed by the All Sky Automated Survey (ASAS), and unveil the Machine-learned ASAS Classification Catalog (MACC), which is a 28-class probabilistic classification catalog of 50,124 ASAS sources.  We estimate that MACC achieves a sub-20\% classification error rate, and demonstrate that the class posterior probabilities are reasonably calibrated.  MACC classifications compare favorably to the classifications of several previous domain-specific ASAS papers and to the ASAS Catalog of Variable Stars, which had classified only 24\% of those sources into one of 12 science classes. The MACC is publicly available  \href{http://www.bigmacc.info}{on-line}.
\end{abstract}

\keywords{methods: data analysis -- methods: statistical -- stars: variables: general -- techniques: photometric -- catalogs}

\section{Introduction}
\label{sec:intro}

Synoptic imaging surveys have begun to routinely collect dozens to thousands of epochs of photometric data over wide swaths of the sky. The manifest destiny for optical time-domain studies is the Large Synoptic Survey Telescope (LSST; \citealt{lsst}), which will collect time histories for $\mathcal{O}$(10$^9$) stars and explosive transients. With data collected for so many sources, no longer is it possible for experts to manually scrutinize significant subsets of the data.  In this era of wide-field time-domain surveys, accurate multi-class source catalogs, which are created automatically by machine-learning (ML) algorithms, are required to maximize the scientific output from these projects (\citealt{2008AIPC.1082..257E,2009AAS...21346029B}).  Furthermore, these catalogs must be \emph{probabilistic} in nature, with well-calibrated posterior class probabilities.  This enables each scientist to use a personalized threshold for selecting objects for follow-up, where science class probabilities fit naturally within a framework for optimizing the allocation of limited resources, and to select objects for population studies, where a rigorous treatment requires detailed understanding of the purities and efficiencies of the sample.

Creating probabilistic multi-class catalogs for large-scale time-domain photometric surveys is a difficult task.  First and foremost, a set of salient class-predictive features\footnote{We define a feature to be deterministic real-numbered or categorical metric based on the time-series input or spatial location of the source. See \S \ref{sec:classification}.} needs to be estimated for each source.  From unevenly sampled light curves which contain seasonal gaps, varying levels of noise, and occasional spurious flux measurements, estimating the periods and amplitudes of oscillation for each source is not trivial.  Furthermore, devising light-curve features that can separate specific sub-classes of sources requires deep domain knowledge.  Next, classification models must be constructed to map the light-curve feature vector for each source to a set of posterior class probabilities.  These classifiers need to be able to automatically learn multiple class boundaries in high-dimensional feature spaces from a set of training data with known classes and, for each source, return a calibrated posterior class probability for each science class.  This endeavor is complicated by the fact that the set of training data is typically not representative of the objects in the survey, which can cause large sample-selection biases (see \citealt{2012ApJ...744..192R}) in the posterior class probability estimates.  Additionally, the sources observed by the survey are not guaranteed to  fit neatly into any of the pre-defined classes, necessitating anomaly detection to identify which sources are likely to belong to an undefined science class.

Several aspects of the catalog effort have required focused research attention.  In \citet{2011rich}, we introduced an end-to-end framework for machine-learned classification of variable stars, with advancements in periodic and non-periodic light-curve feature estimation as well as probabilistic, non-parametric classification methodology.  In terms of classification error rate, our methods showed significant improvement over the previous state-of-the-art (\citealt{2007debo}) on a well-studied data set.  Indeed, other groups have also  converged onto a similar set of tools as the best current light-curve classification methodology for variable stars (e.g., \citealt{2011MNRAS.414.2602D}).  In \citet{2012ApJ...744..192R}, we introduced a methodology to overcome the debilitating effects of non-representative training sets on variable star classification, and in \citet{2012arXiv1201.4863L} we devised methods to appropriately use light curve data from older surveys to classify periodic variable stars in new surveys.  With these advances, the accuracy of variable star classification is improving demonstrably, with cross-validated error rates approaching 15--20\% on multi-class problems with different data sets (\citealt{2011MNRAS.414.2602D,2011rich}).

In this paper, we build on these recent advancements in the photometric classification of variability by focusing on the problem of \emph{how to properly construct a variable star classification catalog from a photometric survey}.  Accurate classification of each source in the survey remains the primary concern of this endeavor.  However, there are several other issues that arise when generating classification catalogs for use in astrophysical studies. First, a classification catalog requires good \emph{calibration} for the posterior class probability estimates, $\Prob(\textrm{class} | \textrm{survey data})$.  Good calibration means that of all the objects for which we estimate a posterior class probability, $p$, of belonging to a certain science class, $p$ proportion of them \emph{truly} belong to that class.  In this paper, we describe a method for calibrating classifier probabilities and outline how such information can and should be used when employing such a probabilistic classification catalog for downstream astrophysical inference.  

Second, when constructing a  classification catalog for a large number of objects, anomalies will certainly be present.  When building a supervised classification model, these anomalies are typically not accounted for, resulting in a classification schema which attempts to cram each object into a predefined classification taxonomy.  In this paper, we describe the use of a semi-supervised anomaly detection routine which allows us to leverage our classifier to determine which sources do not resemble any of the training data and likely belong to a variability class not populated by the training set.  We determine, for each source in the catalog, a real-valued measure of the degree of deviance of that source from the training data.  

Third, in a photometric survey, where a majority of objects will fall near the detection limit, the prevalence of aliased periods (at integer-valued cycles per day for a ground-based survey) with be debilitating for an automated classifier.  We outline, in this paper, a prescription for detecting---and dealing with---sources with aliased periods.  We also detail how to cross-match sources with external catalogs to obtain further classification features (e.g., color) and use a method to impute the values of those attributes when no match is detected.

Finally, we use this methodology to create a calibrated probabilistic classification catalog for a set of 50,124 sources in the All Sky Automated Survey (ASAS; \citealt{1997AcA....47..467P}) based on its publicly available ASAS $V$-band light curve and colors.  Our Machine-learned ASAS Classification Catalog (MACC) contains, for each source, posterior probabilities for 28 different science classes.  This is a wealth of new information compared to the existing ASAS Catalog of Variable Stars (ACVS; \citealt{2002AcA....52..397P}), which classified a subset of these sources into 12 science classes without supplying any posterior class probabilities and giving the uninformative class label `MISC' to a majority of objects.  In addition to probabilistic classifications, MACC gives an anomaly score for each ASAS source, which describes its proximity to objects in the training set.  Furthermore, our catalog provides updated periods, peak-to-peak amplitudes, and dozens of other estimated features for each ASAS light curve.   We ensure that all steps in the MACC catalog creation are transparent and provide a public interface to the catalog at \url{www.bigmacc.info}.  

The paper is structured as follows. In \S\ref{sec:data} we describe the ASAS catalog, our retrieval and pre-processing of the photometric data, and how we cross-match with 2MASS for infrared colors and impute the values of those colors when no match is made.  We describe the classification methodology used to in the construction of the catalog in \S\ref{sec:classification}, including the definition of all light-curve features, description of the period estimation and aliased-period treatment procedures, and the derivation of our labeled training set.  In \S\ref{ss:calib} we describe the procedure to calibrate the classifier posterior probability estimates and in \S\ref{ss:anomscore} we detail the computation of a semi-supervised anomaly score for each source.  The MACC catalog is introduced in \S\ref{sec:catalog}, where we describe the attributes of the catalog and how to access it.  We compare the classifications of MACC to some other ASAS classifications in the literature in \S\ref{sec:compare} and finally we conclude with a few remarks about future directions in \S\ref{sec:conclusions}.

\section{Data}
\label{sec:data}

\subsection{ASAS Data Collection}
\label{ss:asas}

The All Sky Automated Survey\footnote{\url{http://www.astrouw.edu.pl/asas/}}, is an ongoing, long-term project dedicated to the detection and monitoring of the photometric variability of bright stars (\citealt{1997AcA....47..467P}).  Since August 2000, ASAS has monitored bright stars ($V < 14$ mag) in the entire available sky south of $\delta < +28^\circ$ from Las Campanas Observatory.  ASAS uses two small wide-field telescopes to monitor the sky with V- and I-band filters.  Each ASAS telescope takes repeated 180-second exposures using a 2K $\times$ 2K CCD camera with 15-$\mu$m pixels, covering 8.5 $\times$ 8.5 deg$^2$ of the sky (see \citealt{1997AcA....47..467P} for further details).

To date, ASAS has taken more than 267,260 V-band frames, imaging approximately 17 million stars of $V$-band magnitude between 8 and 14.  Of these 17 million objects, ASAS has identified 50,124 variable stars and published the results in the ASAS Catalog of Variable Stars (\citealt{2002AcA....52..397P,2003AcA....53..341P,2004AcA....54..153P,2005AcA....55...97P,acvs}).  The catalog, which contains a rough classification for each source, is made publicly available through the ASAS website, along with $V$-band light curves for 15 million ASAS sources.  For the 50,124 variable stars, ASAS has retrieved a median of 541 usable epochs of $V$-band measurements.  The ACVS light curves range in the number of good detection epochs from 3 to 2232.

\subsection{ASAS Photometric Light Curves}

We retrieved the ASAS ACVS dataset by first referencing the ACVS.1.1 catalog, which contains 50,124 variable stars, and individually retrieving the data for each source from the ACVS website\footnote{available at \url{http://www.astrouw.edu.pl/asas/?page=acvs}.}.  These sources were imported into our DotAstro.org (\url{http://dotastro.org}) astronomical light-curve warehouse for visualization and use with internal frameworks (\citealt{2009ASPC..411..357B}).  Each ASAS source's time-series data file is partioned by its observed field and includes information on the quality of the aperture photometry for each epoch, as well as magnitude measurements (and uncertainties) from up to five different apertures.  Prior to importing the data we chose a single aperture for each epoch using the method detailed below and excluded epochs with a quality \texttt{GRADE=D} or quality \texttt{GRADE=C} when \texttt{MAG=29.999}, which, as detailed in the ASAS data files, corresponds to a non-detection.  Given to the undue influence of extreme photometric outliers in light curve feature estimation, before the generation of time-series based features, we applied sigma clipping to each ASAS light curve, excluding observations that lie beyond 4 standard deviations from each source's mean magnitude.

\begin{figure}
\begin{center}
\includegraphics[angle=0,width=6.5in]{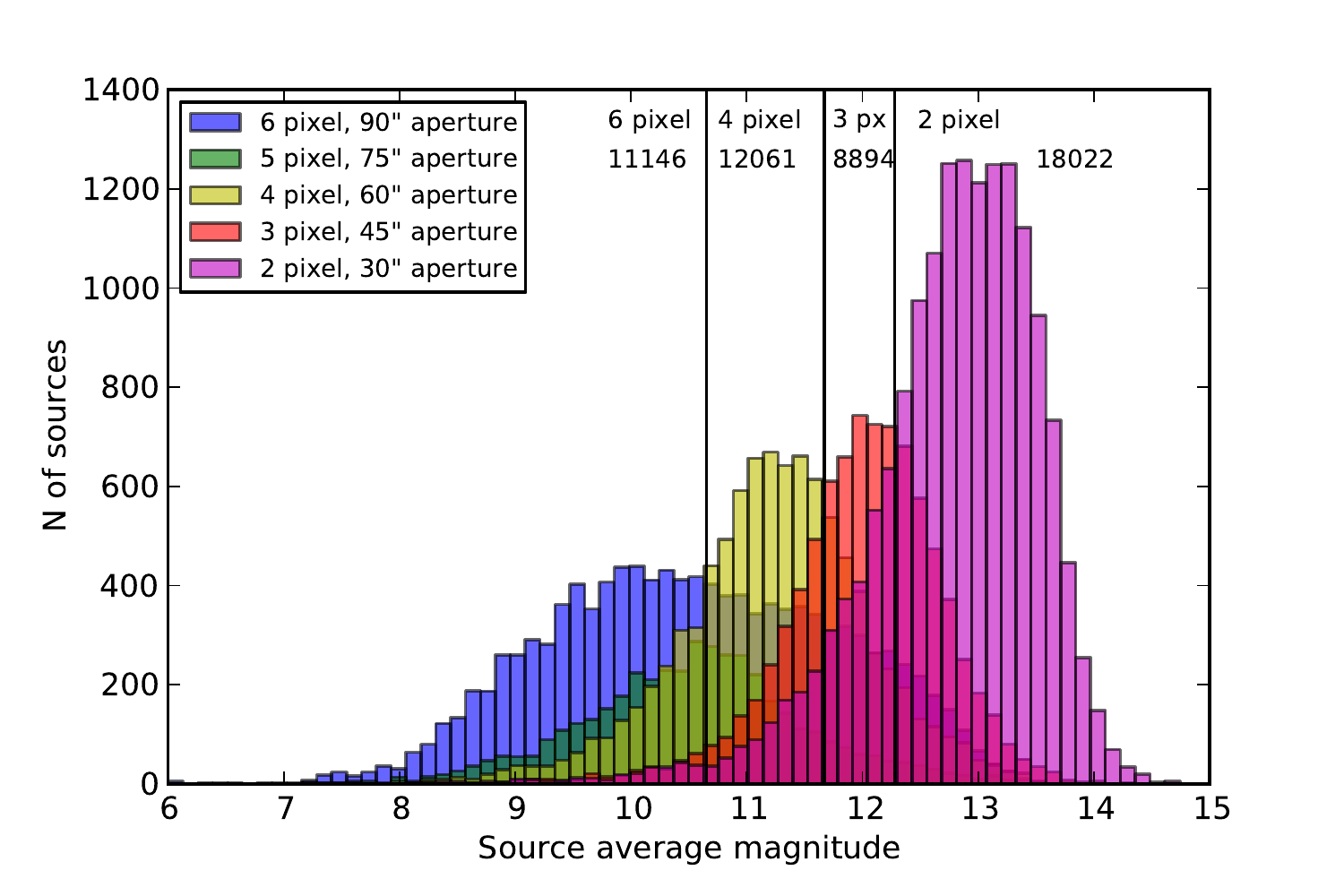}\\
\end{center}
\caption{Aperture-wise histograms of the average magnitude of ASAS sources whose minimal magnitude dispersion was observed to be in the specified aperture.  As expected, brighter sources experience smaller dispersion when observed in wider apertures and fainter sources show smaller dispersion in narrower apertures.  Using these histograms, we construct a kernel density estimation classifier to determine the optimal aperture to use for each ASAS source as a function of its average magnitude of brightness.  The magnitude cuts from this procedure are overlaid in vertical lines and the total number of ASAS objects extracted with each aperture are listed in the figure.
  \label{fig:asas_aperture}}
\end{figure}

ASAS provides five aperture measurements using annuli ranging from 2 pixels (30$^{\prime\prime}$) to 6 pixels (90$^{\prime\prime}$).  Although the ASAS team outlined a basic algorithm for choosing which aperture to use for each source given its average magnitude (\citealt{acvs}), we decided to use our own magnitude-dependent aperture cuts.  Our procedure is the following: we begin by determining the aperture which has the minimum magnitude variance for a source\footnote{Within the field with the greatest number of observations for that source.}.  The idea behind using minimum magnitude variance is that apertures that are too small will not capture all of the source's flux, resulting in larger Poisson noise in the measured brightness of the object, whereas apertures that are too large will incur more background noise and noise due to contamination from neighboring sources.   For each aperture, we can visualize the distribution of mean magnitudes for the sources whose minimal magnitude dispersion occurred in that aperture (see Figure \ref{fig:asas_aperture}).  This information was subsequently employed to construct a simple kernel density estimation (KDE) classifier to determine the optimal aperture to use for each source, as a function of its mean magnitude.  Using this classifier we  determine the optimal aperture  for each ASAS source, as a function of its average magnitude (more precisely, the median magnitude of its five aperture-wise average magnitudes), and only import the light curve measurements from that aperture.  The optimal magnitude cuts for each aperture are over-plotted in Figure \ref{fig:asas_aperture}.

\subsection{Querying the Naval Observatory Merged Astrometric Dataset}
\label{ss:2mass}

In addition to information gleaned from single-band light curves, color information is invaluable to classifying variable stars.
To generate color features, we use the Naval Observatory Merged Astrometric Dataset (\nomad; \citealt{nomad})
 to obtain single-epoch $B, V, R, J, H$, and $K_s$-band photometry for each ASAS object, which we use to compute 5 color features ($B-V, V-R, R-J, J-H, H-K_s$) for each source.  Although the ASAS AVCS catalog provides cross-correlated 2MASS colors, the additional
optical filters provided by \nomad 
supplies a richer set of colors to aid the classifier.  Due to the large ASAS positional errors, we decided against using simple spatial cross-correlation to match each ASAS source
to the \nomad catalog.  Instead, we
train a ML classifier which takes as input 7 positional and photometric features to determine whether a \nomad candidate is indeed a match to the ASAS star.
In addition to the separation distance between the ASAS source and \nomad candidate, we employ the \nomad nearest neighbor rank (ordered by distance from the ASAS source), magnitude differences in $J$, $H$, and $K_s$ bands and $J-K_s$ color difference between ACVS and \nomad, and the $V$-band difference between the ASAS light curve mean magnitude and \nomad to allow a richer view of each source which will facilitate the ASAS--\nomad matching procedure.

This ASAS--\nomad association classifier was initially trained using 48
ASAS sources of known class, sampled from 24 science classes, with 2 sources from
each science class taken from the literature.  For each of these training objects, we manually determined which
source, from a \nomad catalog query around position of the  ASAS source, was truly associated
with that object.  These sources were classified as
`match', while all other sources returned by the \nomad query were
classified `non-match'.

Using the 7 positional and photometric features, we initially trained a random forest (RF, see \citealt{2001brei}) classifier on the 48-object training set
and applied the classifier to predict match/non-match for a  sample of 30,000 of the $\sim$500,000
\nomad sources which are retreived when the \nomad catalog is queried around each of
the 50,124 ASAS sources in our data set.  
Using the active learning technique of \citet{2012ApJ...744..192R}, we selected 17 \nomad sources which would have high impact in improving
the performance of the classifier, and manually classified each as a
`match' or `non-match', and subsequently added these objects to the training set.  This active learning process was performed over 10 iterations, each adding 17  sources, resulting in a
robust classifier which can accurately and automatically decide whether a \nomad source is associated with an
ASAS source based on the positional and photometric features.  

Ultimately, the classification algorithm was applied to each ASAS source to find the matching \nomad entry, if any.   For each ASAS star, we find
 the \nomad source with the highest classifier probability of `match', with preference of spatially closer matches when
identical probabilities are returned for multiple \nomad sources.  If, for an ASAS object, no \nomad source achieves
`match' probability $> 50\%$, then we decide that no \nomad source exists for that object.  When applied
to all 50,124 ASAS sources, we find that 93.9\% of these sources match a \nomad source.

For the 47,044 objects with a \nomad match, we extract the 5 \nomad color features for use in the variable star classifier.  For the remaining 3080 objects with no \nomad match, we impute their colors using the {\tt MissForest} imputation routine of \citet{missforest}.  {\tt MissForest}\footnote{The {\tt R} package {\tt missForest} is freely available at \url{http://cran.r-project.org/web/packages/missForest}.  } is an imputation routine that uses a series of random forests to predict the value of each missing feature based on the observed features for that source.  The {\tt MissForest} algorithm builds a random forest regression model (for real-valued features) or classifier (for categorical features) to predict the value of each feature from all of the other features.  Beginning from some initialization of the missing features, the algorithm iterates until convergence is attained and outputs the predicted value for each missing feature in the data matrix.  On multiple data sets, \citet{missforest} show that {\tt MissForest} outperforms other common methods, such as K-nearest neighbors and Lasso, in imputation accuracy.  We employ {\tt MissForest} using 100 trees.

We test the accuracy of {\tt MissForest} in imputing variable star colors by the following experiment.  Starting with the set of 47,044 objects with a satisfactory \nomad match, we null out the colors for a random 6.1\% of the objects (the same fraction of ASAS objects with no \nomad match).  Then, using the leftover set of sources with known colors, we impute the nulled out colors using {\tt MissForest}.  This allows us to compare the true colors to the imputed colors for this subset of data, which we do using median absolute error (MAE),
\begin{equation}
\sigma(\x_{j, \textrm{imp}}) = \textrm{median}_i \left| x_{ij, \textrm{true}} - x_{ij, \textrm{imp}}   \right|
\end{equation}
where $x_{ij, \textrm{true}}$ and $x_{ij, \textrm{imp}}$  denote the true and imputed values, respectively, of color $j$ for object $i$.  MAEs, $\sigma$, for each of the 5 colors in our data are reproduced in Table \ref{tab:imperr}. While the MAEs for each color, particularly the optical--NIR colors, are larger than the typical uncertainty of the observed color for any individual source, we note that a large scatter is to be expected because we are imputing the observed color without reddening corrections. Indeed, an examination of the observed color for each class shows that the typical within-class scatter is $\gtrsim$ 2 mag, most likely owing to the various galactic latitudes at which the ASAS sources are observed. The imputation procedure confidently identifies stars as being either red or blue, and the obtained accuracy of these imputations is similar to the typical scatter in the observed colors, which gives us confidence that the procedure is sufficient for classification purposes.

\begin{deluxetable}{lcc} 
\tablecolumns{2}
\tablewidth{0pc} 
\tabletypesize{\normalsize}
\tablecaption{Color imputation median absolute errors using the {\tt missForest} imputation method.} 
\tablehead{ 
\colhead{Color} & \colhead{$\sigma$} }
\startdata 
$B-J$ & 0.965\\
$H-K_s$ & 0.059\\
$J-H$ & 0.087\\
$R-J$ & 0.751\\
$V-J$ & 0.863\\
\enddata
\label{tab:imperr}
\end{deluxetable}%

\section{ASAS Variable Star Classifier}
\label{sec:classification}

Probabilistic supervised light curve classification has recently received much attention in the literature.  For example, \citet{2007debo}, \citet{2011MNRAS.414.2602D}, and \citet{2011rich} have applied modern machine learning methods to ${\sim} 25$-class variable star problems using photometric light curve data from the \hipp and OGLE surveys.  This automated classification methodology consists of the following two-step process:
\begin{enumerate}
\item From each light curve, a set of $m$ \emph{features} (e.g., period, amplitude, etc.) is extracted.  These features are constructed to capture the class-predictive information encoded within each light curve.
\item Using a training set of objects of known class, a classification model, which maps from $m$-dimensional feature space to the set of classes, is fit.  Methods such as neural nets, decision trees, support vector machines and random forests are classification models that have been used for light curve classification.  The fitted classification model serves as a class prediction engine.
\end{enumerate}
 Once the classifier has been trained, it is trivial to predict the class of each variable star, which entails first extracting the feature vector of the object and subsequently inserting that vector into the classifier to obtain a prediction.  Many classifiers, such as random forest, produce a vector of posterior class probabilities for each object.
 
 To construct the ASAS variable star classification catalog, we use a set of $m$ = 71 features:  66 light curve features and 5  colors (described in \S\ref{ss:2mass}).  See \S\ref{ss:feats} below for a description of the features used.  We use a random forest classifier, which has been shown to attain high levels of accuracy in variable star classification by \citet{2011MNRAS.414.2602D}, and \citet{2011rich}.  \citet{2011rich} found that the random forest classifier attained the lowest error rates in classifying \hipp and OGLE variable stars in a side-by-side comparison with a dozen other classification models.  In \S\ref{ss:train} we describe how to attain a training set for ASAS to minimize classification errors due to sample selection bias (see \citealt{2012ApJ...744..192R} for a thorough discussion of sample selection bias for light curve classification).

\subsection{Light Curve Feature Extraction}
\label{ss:feats}

Raw light curve data consist of measurements of a source's brightness over unevenly sampled epochs.  From these data, our challenge is to estimate a set of features that are predictive of each source's class (e.g., it is well known that period, amplitude, and color are all highly predictive of class for certain classes of  pulsating variable star) while being agnostic to other latent factors which are unrelated to (or at most, mildly correlated with) an object's science class.  Examples of such latent factors are that each ASAS light curve consists of a different number of epochs (ranging from 3 to 2232 epochs with median of 541), over a different time baseline, with distinct noise properties and differing cadences.  Furthermore, each ASAS source has a unique mean brightness (from 4th to 15th magnitude in V), resides in a unique position in the sky, and has its light affected by more or less intervening dust.

We have constructed a set of 66 light-curve features meant to capture the essence of photometric variability of the science classes of interest, and have written algorithms that efficiently compute these features from light curve data, in an average of 4.5 seconds per ASAS light curve.  In \citet{2011rich}, a set of 52 features was used to represent each variable star.  Below, we describe the additional features that have been used in this study, and also outline some modifications to the algorithms used for periodic modeling.

\subsubsection{Computationally Efficient Regularized Fitting of Periodic Signals of Arbitrary Shape}

In this study, we employ a novel fitting routine which seeks to simultaneously
discover the true period of a source while also modeling the light
curve in detail.

We begin by applying our fast Lomb-Scargle algorithm (fit of single
sinusoid; \citealt{2011rich})
to discover all marginally significant periods for a given light curve
on a broad frequency
test grid ($\nu_{\rm min}=1/T$, $\nu_{\rm max}=10$, $\delta \nu =
0.1/T$ cycle/day, where $T$ is
the data timespan).  For test frequencies where the power-spectrum has
a value $>6$ (i.e., $< $1\%
of test points, corresponding to roughly $3.5\sigma$ significance), we
fit a multi-harmonic model,
\begin{equation}
\label{eqn:lombscargle}
m_i = ct_i +  \sum_{n=1}^{8} A_n \sin(2\pi \nu_0 n t_i) + B_n \cos(2\pi \nu_0 n t_i) + b_{n,o}
\end{equation}
consisting of a sinusoid at the initial frequency, $\nu_0$, plus sinusoids at
each of the $n=(2,...,8)$ harmonics of that initial frequency.     We choose $n=8$ to allow for sufficient
model complexity to account
for the light curves under study.  The fitting of model \ref{eqn:lombscargle} is performed with a
regularization penalty to avoid
over-fitting, and the number of effective model degrees of freedom is
typically well below the allowed value
of $2\times 8=16$.

In the fitting, we minimize
\begin{equation}
 R = \sum_{i=1}^N \frac{(d_i-m_i)^2}{\sigma_i^2} + N\lambda* \sum_{n=1}^8 n^4 (A_n^2+B_n^2),
\end{equation}
with respect to the model parameters $\theta$ and the regularization
parameter $\lambda$.
Here, the photometric data are $d_i$, the model is $m_i$, $N$ is the number of data points,
and $\sqrt{A_n^2+B_n^2}$ is the amplitude of the $n$th Fourier
harmonic.  The second
term above effectively penalizes the model in proportion to the
magnitude of its second derivative.
Small values of $\lambda$ result in models with high-frequency
structure, whereas large $\lambda$
values yield more smooth, slowly changing models.
For fixed $\lambda$, the best fit parameters can be found by least-squares.  We
identify the optimal value of $\lambda$ using generalized cross
validation (\citealt{golub1979generalized,craven1979smoothing}).
This allows the data to drive the complexity of the model while also
constraining the model to
not over-fit the data.

\subsubsection{Novel Light-Curve Features}

In addition to the 32 periodic and 20 non-periodic features used in \citet{2011rich} to parametrize variable stars, we add 15 new features based on our generalized Lomb-Scargle periodogram, of which ten were also used by \citet{2012arXiv1201.4863L}.  These features are compiled in Table \ref{tab:newfeats}.  The first two features are \verb freq_amplitude_ratio_21  and \verb freq_amplitude_ratio_31 , which are ratios of the amplitudes of the the second to first and third to first frequencies, respectively.  We also add three features aimed at detecting eclipsing sources from the  Lomb-Scargle model in Equation (\ref{eqn:lombscargle}), phased on twice the Lomb-Scargle period.  We compute the phases and magnitudes of the two distinct minima and two distinct maxima of the phased light curve model.  The feature  \verb freq_model_max_delta_mags  is the absolute value in the magnitude difference between the two model light curve magnitude maxima (i.e. eclipses), and should be non-zero if the source is an eclipsing binary.  Similarly, the feature \verb freq_model_min_delta_mags  captures the absolute value in the magnitude difference between the two magnitude minima and the feature \verb freq_model_phi1_phi2 , which is constructed to detect eccentric binary systems, is the ratio of the phase difference between the first minimum and the first maximum (i.e. primary eclipse) to the phase difference between the first minimum and second maximum (i.e. secondary eclipse).

Additionally, we introduce the feature \verb freq_n_alias , which counts the number of frequency estimates that are consistent with a 1-day alias (see \S\ref{ss:alias} for details of this procedure).  This feature supplements the \verb freq_signif  feature to determine whether a source is, in fact, periodic.  A source with estimated period at a 1-day alias often has a large \verb freq_signif  value even when the light curve is truly aperiodic, in which case it will be identifiable as aperiodic by a non-zero value of \verb freq_n_alias .  We further add the class-specific feature \verb freq_rrd , which indicates whether any of the frequency ratios are consistent with 0.746, which is the frequency ratio enjoyed by double-mode RR Lyrae variable stars.

Finally, we add the following 5 features which are adopted from  \citet{2011MNRAS.414.2602D}.  The feature \verb scatter_res_raw  computes the ratio of the median absolute deviation (MAD) of the residuals of the Lomb-Scargle model to the MAD of the raw light-curve.  The features  \verb p2p_scatter_2praw , \verb p2p_scatter_over_mad  , and \verb p2p_scatter_pfold_over_mad  are the sum of squared differences of the scatter about the light curve phased on the Lomb-Scargle period to that of either the phased or raw light curve data.  Similarly, the feature  \verb medperc90_2p_p  is the 90th percentile of the absolute residual values around model phased on twice the Lomb-Scargle period divided by the same quantity for the residuals around the model phased on the Lomb-Scargle period.  Furthermore, we develop two new features, \verb fold2P_slope_10percentile , and  \verb fold2P_slope_90percentile , which are the 10th and 90th percentile slopes of the Lomb-Scargle model around twice the period, intended to capture the steepness of the ingress and egress of eclipse.  Lastly, we add the feature \verb p2p_ssqr_diff_over_var  from \citet{2011arXiv1101.3316K}, which is the sum of squared magnitude differences in successive measurements divided by the variance.

\begin{deluxetable}{ll}
\tablecolumns{2}
\rotate
\tablewidth{0pc} 
\tabletypesize{\footnotesize}
\tablecaption{Light-curve features used in addition to the features of \citet{2011rich}.} 
\tablehead{ 
\colhead{Feature} & \colhead{Description}}
\startdata 
\verb freq_amplitude_ratio_21  & amplitude ratio of the 2nd to 1st Fourier component in the Lomb-Scargle model\\
\verb freq_amplitude_ratio_31  &  amplitude ratio of the 3rd to 1st Fourier component in the Lomb-Scargle model \\
\verb freq_model_max_delta_mags  & absolute value of mag difference between the two model light curve  maxima phased on 2P\tablenotemark{a}  \\
\verb freq_model_min_delta_mags  & absolute value of mag difference between the two model light curve  minima phased on 2P \\
\verb freq_model_phi1_phi2  & ratio of the phase difference between the first minimum and the first maximum to the phase \\
&  difference between the first minimum and second maximum \\
\verb freq_n_alias  & number of top period estimates that are consistent with an alias \\
\verb freq_rrd  & boolean that is true only if \verb freq_frequency_ratio_21  or \verb freq_frequency_ratio_31  \\
& are consistent with 0.746\\
\verb scatter_res_raw  & MAD of the Lomb-Scargle residuals divided by the MAD of the raw light-curve values \\
\verb p2p_scatter_2praw  & sum of squared mag differences between pairs of successive observations in the light curve folded \\
&  around 2P divided by that of the raw light curve \\
\verb p2p_scatter_over_mad  & median of the absolute differences between successive observations normalized by the MAD \\
\verb p2p_scatter_pfold_over_mad  & median of the absolute differences between successive mags in the folded light curve normalized by \\
& the MAD of the raw light curve\\
\verb medperc90_2p_p  &  90th percentile of the absolute residual values around the 2P model divided by the same \\
& quantity for the residuals around the P model\\
\verb fold2P_slope_10percentile  & 10th percentile of slopes between adjacent mags after the light curve is folded on 2P \\
\verb fold2P_slope_90percentile  & 90th percentile of slopes between adjacent mags after the light curve is folded on 2P \\
\verb p2p_ssqr_diff_over_var  & the sum of squared mag differences in successive measurements divided by the variance\\
\enddata
\tablenotetext{a}{We use P to denote the Lomb-Scargle estimated period, and 2P to be double that period.}
\label{tab:newfeats}
\end{deluxetable}%

\subsubsection{Correcting Eclipsing Periods}

Comparison of our estimated periods with those from the ACVS catalog reveals that our period estimates are often exactly half of the ACVS period for sources which are classified as eclipsing binaries by ACVS.  Of the 5913 objects that are classified as eclipsing binaries in ACVS, our period estimate matches the ACVS period for only 1339 sources (23\%) and was exactly one-half of the ACVS period for 4162 sources (70\%).  After visual inspection of some of these light curves, we find that for eclipsing binaries in which our periods differ, the ACVS period is correct for most (but not all) of the objects.  Using a visually confirmed set of 150 eclipsing sources in which our period is exactly one-half of the true period and 150 eclipsing binaries for which our period is correct, we construct a supervised machine-learned random forest classifier on all of the features described in (\S\ref{ss:feats}) to automatically discover, for each eclipsing source in the data set, whether our estimated period is correct or wrong by a factor of one-half.

In this classifier, the most important features in determining whether our period is correct are, unsurprisingly, the \verb freq_model_max_delta_mags , \verb freq_model_min_delta_mags , and \verb freq_model_phi1_phi2  features, which capture differences between the primary and secondary eclipses, and \verb freq1_harmonics_amplitude_1 ,  the amplitude of the first harmonic, which will be large for an eclipsing binary containing two unequal eclipse depths that was incorrectly identified as having period one-half of the true eclipsing period.  We apply this classifier to all 11,169 sources in our data set that were either classified by ACVS as an eclipsing binary or whose most probable class from our variable star classifier was one of the eclipsing binary classes.  Of those sources, the classifier determined that our period was correct for 5807 objects and that our period was wrong by a factor of 1/2 for 5516 sources.  Doubling the period of those 5299 sources yielded a significant boost in the period agreement rate with the ACVS eclipsing binary stars, with 4225 of 5913 (71\%) of those sources resulting in a period match.

\begin{figure}
\begin{center}
\includegraphics[angle=0,width=5in]{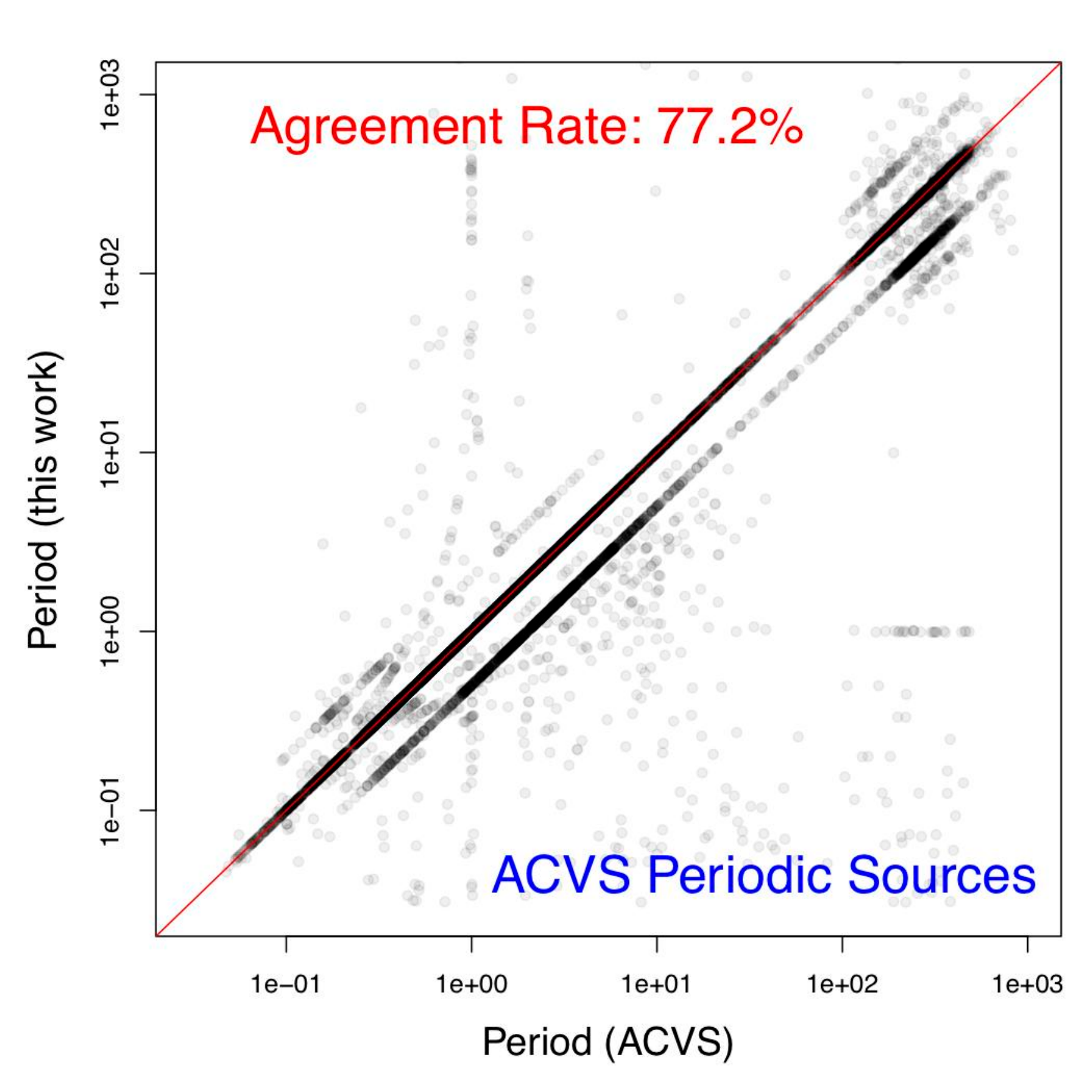}
\end{center}
\caption{ Period estimated by our period-finding algorithm versus the period stated in the ACVS catalog, for all 12,008 ASAS periodic sources in ACVS.  The red dashed line denotes perfect agreement; for a total of 9280 of the stars (77.2\%) we find periods that exactly match the ACVS period.  For 92.9\% of these sources, our period estimate either matches the ACVS period exactly or is different by a factor of two. \label{fig:periodplot} }\end{figure}

In Figure \ref{fig:periodplot} we plot, for the 12,008 ASAS sources which the ACVS confidently classified into a single periodic class (i.e., not classified as ``MISC" and not listed in multiple classes), the ACVS period versus our estimated period.  Our agreement rate with ACVS is 77.2\% on these objects.  Including matches to half and twice the ACVS period yields an agreement rate of 92.9\%.  For 14.1\% of the ACVS periodic sources, our period finder estimates a best-fit period of exactly half the ACVS period and for 1.6\% of the sources our period is double the ACVS period.   Though some of these cases may be errors on our part, they are not debilitating to the variable star classifier because 1) eclipsing binaries are not constrained to inhabit a narrow period range, and 2) other eclipsing binary features are more useful in detecting binary systems and sub-classifying them into their physical class.

\subsubsection{Treating Aliased Periods}
\label{ss:alias}

\begin{figure}
\begin{center}
\vspace{-0.35in}
\includegraphics[angle=0,width=4in]{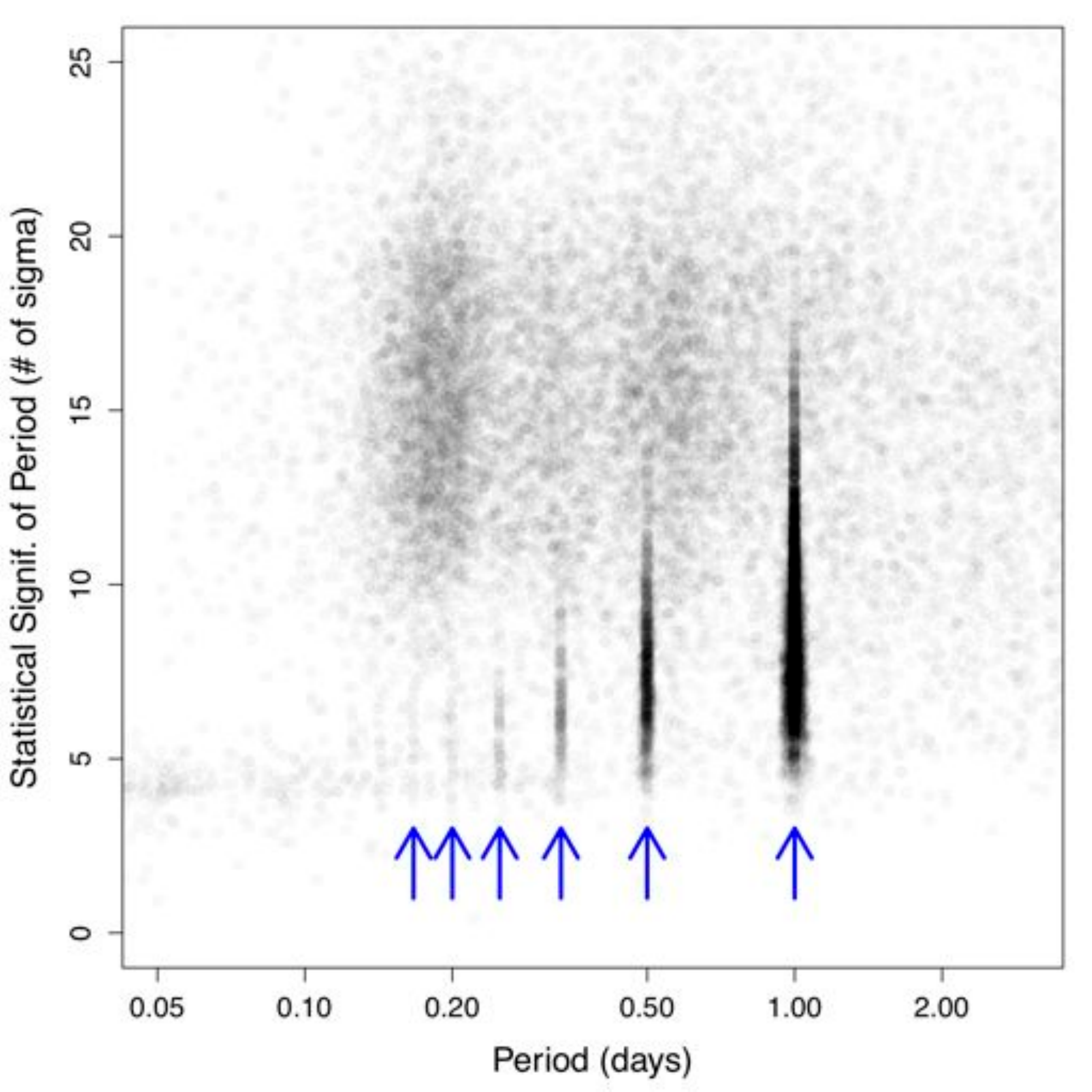} \\ \vspace{-0.05in}
\includegraphics[angle=0,width=4in]{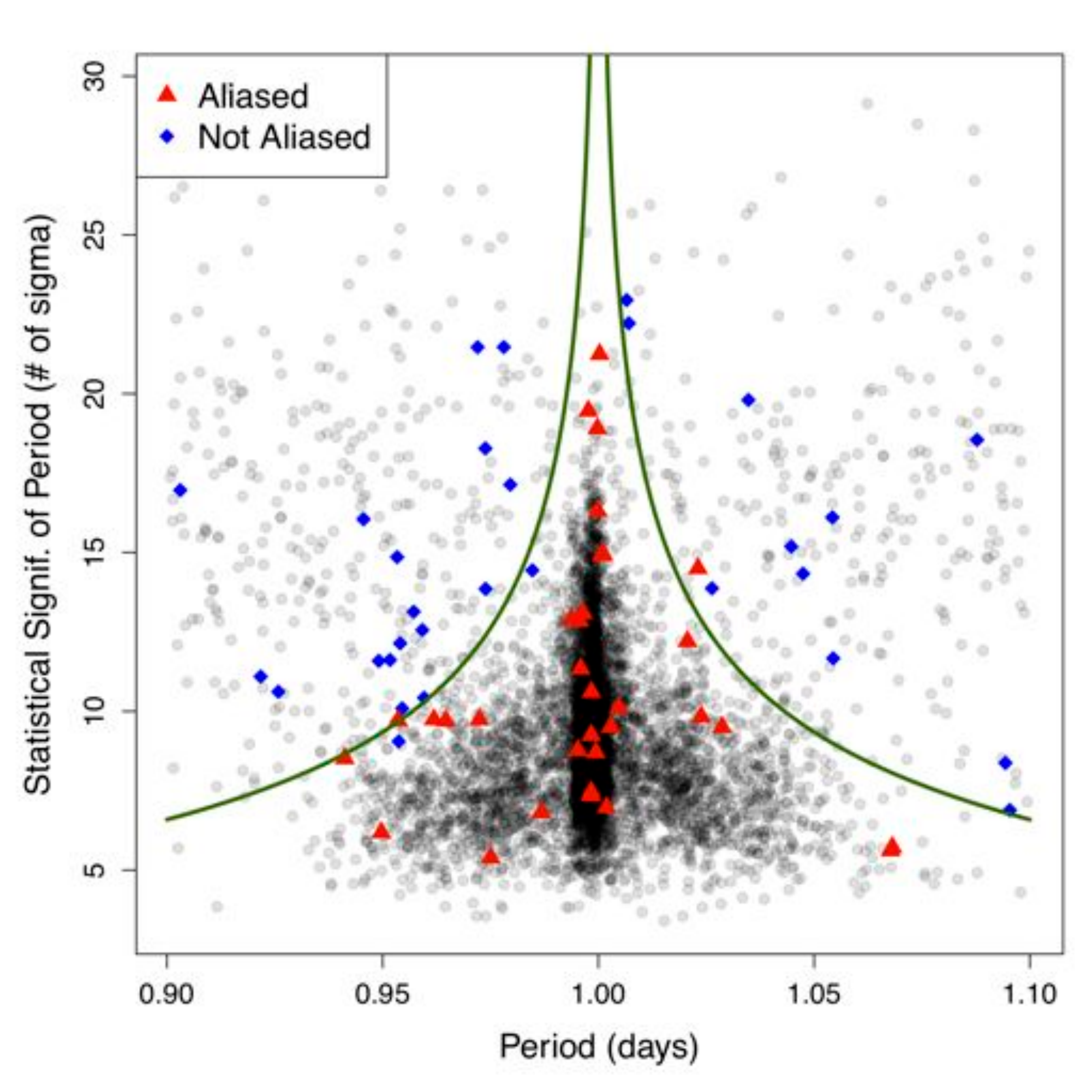} \vspace{-0.3in}
\end{center}
\caption{Top: Period versus period significance (in number of $\sigma$) for ACVS objects, estimated from their ASAS $V$-band light curves.  There are clear over-densities at 1-day, 1/2-day, etc. periods (as denoted by the arrows).  The aliased objects typically have small statistical significance in the period.  Bottom: Zooming into the region around 1-day periods, we see a clear division between objects with aliased (red $\blacktriangle$) and non-aliased (blue $\diamond$) periods, as verified by manual study.  We find that a function of the form of Equation \ref{eqn:alias} separates the aliased and non-aliased objects; we use this function to decide whether each source is aliased at a 1-day period, and similarly for other aliases.  \label{fig:alias}}\end{figure}

In a ground-based survey, aliases are common at 1-day periods due to the rotational period of the Earth.  Large samples from photometric surveys are generally filled with quasi-periodic and non-periodic sources and low S/N periodic light curves, making 1-day aliases prominent.  The prevalence of 1-day aliases  in ASAS, as well as aliases at each integer number of cycles per day, is easily seen in a plot of estimated period versus statistical significance of the period (Figure \ref{fig:alias}).  There is a clear division between aliased and non-aliased sources, as aliases objects tend to have smaller statistical significance, as exhibited by the over-density of objects around each aliased period with significance $\lesssim 10 \sigma$.

We use the period--period-significance plane to determine whether each source's period is an alias.  At each period $P = 1, 1/2, 1/3, 1/4$ days we perform the following experiment.  First, we randomly sample 25 objects from a small window around the period, $P$.  Using the {\tt ALLSTARS} web-based visualization tool\footnote{Active Learning Lightcurve classification Service, see \citet{2012ApJ...744..192R} for more details.}, we decide whether each object's period is truly aliased.  Next, we sample additional objects whose determination of alias versus non-alias  is not clear from the initial sample; this is achieved by fitting a 5-nearest neighbor classifier to the initial set of 25 objects and only selecting objects whose 5 nearest neighbors disagree in a 3-to-2 ratio.  These objects are again verified manually and the process repeated until the phase space is appropriately filled in.  Finally, we find that a function of the form
\begin{equation}
s_P(x_i) = \frac{\alpha_{1,P}}{ | x_i - P | ^{1/4}} + \alpha_{2,P}
\label{eqn:alias}
\end{equation}
produces an acceptable separation boundary between the aliases and non-aliases around each period $P= 1, 1/2, 1/3, 1/4$, where $x_i$ is the estimated period for object $i$ and $s_P(x_i)$ is significance below which the object is deemed to be an alias.  Using the above sample, we find the values of $(\alpha_{1,P}, \alpha_{2,P})$ that minimize the sum of the squared residual distances for objects whose alias-ness is misclassified.  In Figure \ref{fig:alias}, we plot the estimated alias-decision boundary for 1-day aliases, along with the training sample used to determine that boundary.  This allows us to determine, for any object near an aliased period, whether it is aliased, and enables us to select its next-best non-aliased period as the true period and to compute the afore-mentioned \verb freq_n_alias  feature.

\subsubsection{Feature Importance}
\label{sss:featimp}

\begin{figure}
\begin{center}
\includegraphics[angle=0,width=6in]{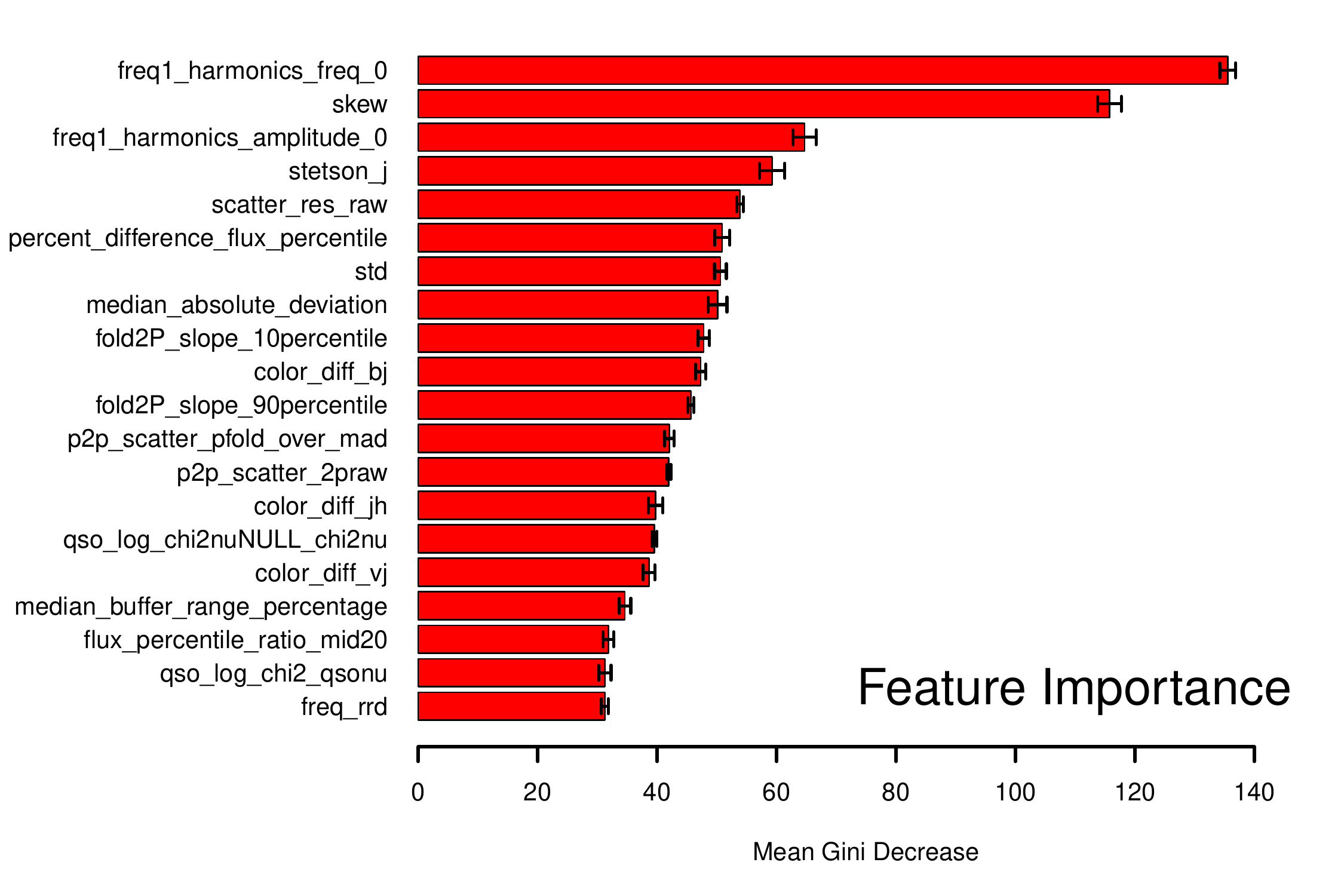}
\end{center}
\caption{ Random forest feature importance for the top 20 features, as estimated by calculating the mean feature importance over 5 random forest classifiers.  As expected, the fundamental frequency of oscillation is the most important feature in ASAS variable star classification.  The next most important features include the skew of the flux measurements, the Fourier model amplitude of the fundamental frequency, the $B - J$  color, the ratio of the standard deviation of the scatter about the Fourier model to the raw observed scatter, and the Stetson variability index $J$ (\citealt{1996stet}).  Error bars denote the standard deviation in the feature importance over 5 random forests. \label{fig:featimp} }\end{figure}

For completeness, we plot the importance of each feature in the classification random forest in Figure \ref{fig:featimp}.  The RF feature importance measure describes the decrease in overall classification accuracy that would result if the feature were replaced by a random permutation of its values.  See \cite{2001brei} for further details.  In Figure \ref{fig:featimp} we find that the fundamental frequency of oscillation (i.e. period) of the light curve is by far the most important feature in the classifier.  Other important features include estimates of the light curve skew, measurements of amplitude/variability (\verb stetson_j , \verb std , \verb median_absolute_deviation), various colors, and features extracted from the light curve folded on twice the period.  One caveat to the feature importance measure is that it does not account for correlations between features.  For instance, the standard deviation and median absolute deviation of the light curve both provide measurements of the spread in the flux measurements about the average value; thus, the conditional importance of \verb std  given \verb median_absolute_deviation  is quite low even though their individual importance measures are both large \footnote{\citet{2011MNRAS.414.2602D} account for this by iteratively removing features that are highly correlated with the most important features.  We choose not to perform this preselection because we find that some redundancy in the features helps the performance of random forest classifier, especially when using features that are robust to outliers.}.

\subsection{Training the Classifier}
\label{ss:train}

Non-parametric supervised classification methods, such as random forest, require a training set of data with known class label to learn the mapping from feature space to classes.  Once this model is learned, data from each ASAS source can be trivially fed into the model to attain probabilistic classifications for each object.  However, much care must be taken to attain a training set that is representative of the ASAS data.  If significant discrepancies exist between the distribution of training features and the distribution of the features of the ASAS data, then, as shown by \citet{2012ApJ...744..192R}, significant biases can occur in the ASAS classifications due to poor model selection and catastrophic errors caused by sample-selection bias.  In this section, we detail the construction of our classification training set and efforts to avoid sample-selection bias.

As the base training set for the ASAS classifier, we use the training set of confirmed \hipp and OGLE sources used in \citet{2011rich} (which is based on, but slightly different than the training set used by \citealt{2007debo}).  This data set consists of 1549 variable stars from 27 different science classes\footnote{Note that this training set is slightly different than that of \citet{2011rich} in that we further split the T Tauri class into Classical (9 stars) and Weak-line (2 stars) subclasses and add  the SX Phoenicis variable class.}.  Next, we cross-match the \hipp training set with our ASAS sample, finding 268 matching sources.  For these 268 training objects, we replace their \hipp light curves with their ASAS light curves in the training set.  At this stage of the analysis, we also choose to exclude four variable star classes: Lambda B\"ootis, Slowly Pulsating B, Gamma Doradus, and Wolf-Rayet.  Each of these classes of variable star is populated by objects whose amplitude of variability is $\Delta V \lesssim 0.05$ mag, which is below the ACVS variability selection cut of 95th percentile in the magnitude-dispersion diagram (\citealt{2002AcA....52..397P}).  Indeed, of the 113 variable stars in our \hipp training set that belong to one these four classes, not a single star passed the variability cuts used to construct the ACVS catalog, even though 78 of the 113 stars were observed by ASAS.  Because such prototypical examples of each of the four small-amplitude classes did not satisfy the cuts used to construct ACVS, we do not expect to find any objects of these classes in the ACVS sample.

The feature distribution of this initial training set is substantially different than the bulk distribution of ASAS features (see Figure 1 of \citealt{2012ApJ...744..192R}).  In \citet{2012ApJ...744..192R} it was exemplified that this mismatch causes poor performance by supervised ML classification and demonstrated that an active learning framework could be used to supplement the training set in a statistically rigorous manner.  Active learning is a classification paradigm in which the supervised classifier is able to query the human user for the classification labels of a subset of sources with unknown class, whereby these objects are manually labeled by the user and added to the training set.  Using a random forest classifier, the active learning query function $S_2$ from equation (5) of \citet{2012ApJ...744..192R}, and the crowd sourcing methodology outlined in that work, we add 407 ASAS sources to the training set. 

In addition to the 407 active-learning training sources, we supplemented the classification training set
with matched sources from the SIMBAD catalog (\citealt{simbad}) using a combination of algorithmic
catalog matching, literature searching, and human vetting.  Starting
with the list of \nomad sources associated with ASAS sources (see \S\ref{ss:2mass}), our algorithm looks
for a SIMBAD source which is spatially close to the \nomad source, calling a match any SIMBAD source which is within 0.5 arcseconds
of the \nomad source.  If no SIMBAD source fits this constraint, then 
no association is made.  Our primary purpose for this exercise was to strengthen the training set for under-represented science classes.  Thus,
 for any positive SIMBAD association of class RV Tauri, Population II Cepheid, Beta Cephei, Chemically Peculiar, T Tauri, or Herbig Ae/Be,
 we performed a literature search on the object, only including the source in the training set if it was definitely confirmed by multiple sources.    This procedure
 allowed us to add 68 sources to the training set.  At this point, we also added R Coronae Borealis (RCB)---a well-studied class of hydrogen-deficient carbon-rich supergiants that undergo episodes of extreme dimming (\citealt{1996PASP..108..225C})---to the training set, populating the training sample with 17 RCB stars found via the SIMBAD matching procedure.

In a preliminary edition of the classification catalog it was noticed that an excessively large fraction 
of the ACVS variables, $\gtrsim 10$\%, were being classified as T Tauri stars (TTS). At the time TTS only 
constituted $\sim$0.7\% of the training set so the large fraction of TTS classifications was not expected.
Upon further inspection we discovered that the inclusion of the two sub-classes of TTS, which exhibit 
significantly different photometric behavior, into a single class led to their significant overrepresentation 
in the final catalog. Thus, we decided to split the TTS class into two classes: weak-line T Tauri stars (WTTS) 
and classical T Tauri stars (CTTS). This split is physically motivated as WTTS are older young stellar objects 
whose photometric variability is periodic and characterized by the rotational modulation of cool spots on the 
stellar surface; CTTS, on the other hand, are younger stars that are still actively accreting from a disk with 
a variability signature that is typically more chaotic than WTTS (for a review of TTS variability see \citealt{Herbst94} 
and references therein). To populate these two new classes we divided all members of the original TTS training 
set as well as new TTS identified via our SIMBAD--ASAS matching query, which included SIMBAD matches of type Y*O, Or*, pr*, 
or TT*\footnote{Y*O: Young Stellar Object; Or*: Variable Star of Orion Type; pr*: Pre-main sequence Star; TT*: 
T Tau-type Star.}. We split these sources into the CTTS and WTTS classes using the classical diving line between the two: 
for CTTS the equivalent width (EW) of H$\alpha$ emission is $>$10 \AA, while for WTTS EW$_{{\rm H}\alpha} < $10 \AA\ 
(see e.g., \citealt{Walter86,Strom89}). Stars were only included in the training set if we could find a published value 
of EW$_{{\rm H}\alpha}$, which typically came from the catalogs of \citet{herbig88} or \citet{Torres06}.

It was later noticed that several known members of the RS Canum Venaticorum (RS CVn) class of binary stars were being 
classified as WTTS, which prompted us to add RS CVn stars as a new class in the training set.  To populate the RS 
Canum Venaticorum class in the training set we identified matches between ACVS sources and the catalog of 
chromospherically active binary stars (CABS; \citealt{Strassmeier88}). The CABS includes both RS CVn and BY Draconis 
(BY Dra) binaries, both of which we include in the training set as the latter is the low mass analog of the former. In 
practice RS CVn and BY Dra stars exhibit the same photometric behavior, from a classification standpoint they can only 
be separated spectroscopically which is why we include them as a single class in the MACC. The cross-match between the CABS 
and ACVS produces 16 RS CVn and 1 BY Dra which we use to define the RS CVn training set.

Our final training set consists of 1945 sources in 28 science classes. A total of 777 of these sources are observed by ASAS, so we use their ASAS light curves to derive features to train the classifier on.  For the other 1168 training objects, we only have data in \hipp (644 stars) or OGLE (524 stars), so we employ the light curves observed by those missions.  A tabulation of the entire training set, by class, is given in Table \ref{tab:training}.  The implicit class prior in fitting a random forest classifier is the empirical vector of training-set class proportions, which is given in Table \ref{tab:training}.

Finally, we find the optimal random forest model by minimizing the 10-fold cross validation classification error rate over the ASAS training set with respect to the number of random forest trees, {\tt ntree}, the number of features considered on each splitting node, {\tt mtry}, and the minimum size of each terminal node, {\tt nodesize}.  Performing a grid search over those three parameters, we find that the optimal model is {\tt ntree} $= 5000$, {\tt mtry} $= 17$, and  {\tt nodesize} $= 1$, attaining an average 10-fold cross validation error rate of 19.15\% for the 777 ASAS training objects.  For the remainder of this paper, and to construct the ASAS classification catalog, we use this optimized classification model.

\begin{deluxetable}{lcc} 
\tablecolumns{3}
\tablewidth{0pc} 
\tabletypesize{\footnotesize}
\tablecaption{Class distribution of training set objects used to fit the probabilistic ASAS classifier.  This class distribution defines the prior on class probabilities used to compute posterior class probabilities for each source.} 
\tablehead{ 
\colhead{Science Class} & \colhead{$N_{\rm Train}$} & Prior $\Prob$(Class)}
\startdata 
\input{training_set.dat}
\enddata
\label{tab:training}
\end{deluxetable}%

\subsection{Calibrating Classifier Probabilities}
\label{ss:calib}

Using the features described in \S \ref{ss:feats} and the training set outlined in \S \ref{ss:train}, we fit a random forest classifier with optimized tuning parameters and use it to generate class predictions and full 28-class probability vectors for all 50,124 ASAS objects.  A desirable property of probabilistic classifications is that they be \emph{calibrated}.  That is to say, if we consider all sources whose class probabilities for a particular class are 90\%, then 90\% of those objects should truly be of that class.  Calibration is attractive because it allows us to treat the probabilistic classifier output as if it were truly a set of posterior class probabilities, $\Prob(\textrm{class} \,|\, \x)$.  Calibration also allows us to easily substitute different prior class probabilities by multiplying the classification probabilities by the appropriate vector of prior ratios and re-normalizing the probability vectors (see \ref{ss:priors} for a detailed explanation).

\begin{figure}
\begin{center}
\includegraphics[angle=0,width=6in]{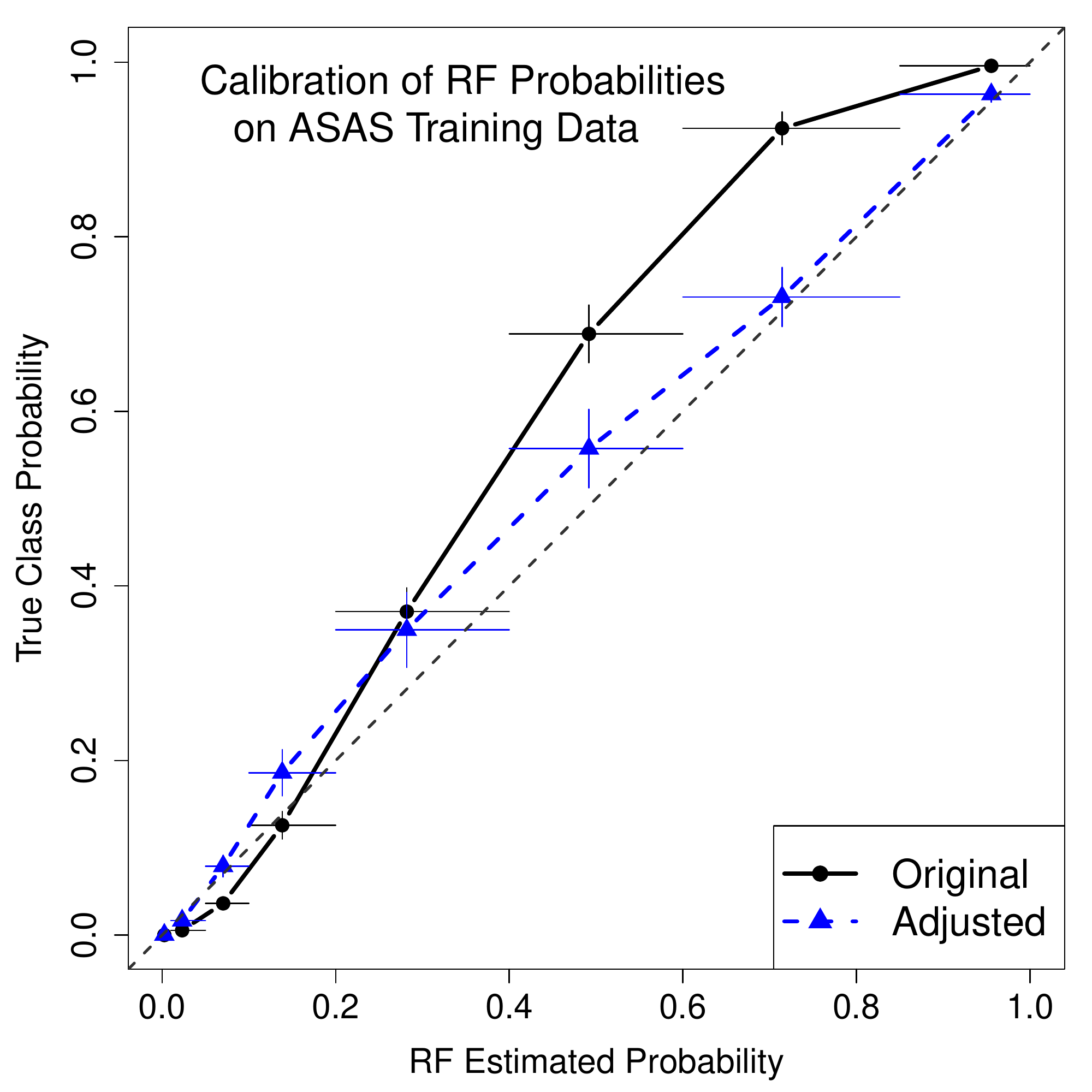}
\end{center}
\caption{ Reliability diagram for ASAS training data.  The closer that the curve follows the diagonal, the better calibrated that the classifier probabilities are.  The initial random forest probabilities (solid black line) are not well calibrated, as the RF probabilities tend to grossly underestimate the true posterior probabilities for large estimated probabilities.  Using the calibration procedure of \citet{bostrom2008calibrating} results in well-calibrated adjusted probabilities (dashed blue line) as they are consistent with the diagonal of the reliability diagram.  In the final ASAS catalog, we use this calibration procedure to adjust all of the posterior probability estimates. \label{fig:calibration} }\end{figure}

However, the class probabilities estimated by the RF are not necessarily calibrated.  To check their calibration we perform the following experiment.  Using only the subset of ASAS training data (777 objects), we perform 10-fold cross-validation to estimate the random forest classification probabilities for each source\footnote{Cross-validation ensures that each object is held out of the training set when fitting the classifier that is used to predict the class probabilities for that object.  In this sense, the cross-validated classification probabilities are representative of the classifier probabilities for the unlabeled data.}.  This provides a vector of 28 cross-validated class probabilities for each object.  Then, in each of 8 disjoint probability bins (chosen such that each bin contains at least 100 instances), we compute the proportion of the objects, $p_{\rm true}$, that are truly of the specified class.  If the probabilities were calibrated, then the value of $p_{\rm true}$ should match the mean random forest probability within each bin.  In Figure \ref{fig:calibration} we see, by the solid black line, that this certainly is not the case for our classifier.  Specifically, the random forest classifier tends to be \emph{conservative} in that it systematically estimates a smaller probability than $p_{\rm true}$ for RF probabilities greater than ${\sim}0.3$.  For instance, in the RF probability bin centered around 0.5, around 70\% of those objects are truly of the specified class.

Two popular methods exist for calibrating classifier probabilities using simple transformations.  
\emph{Platt Scaling} (\citealt{platt1999probabilistic}) transforms the probabilities using a sigmoid function whose parameters are chosen via maximum likelihood over the training set.  \emph{Isotonic Regression} (\citealt{robertson1988order,zadrozny2001obtaining}) is more flexible, replacing the sigmoid function with any monotonically increasing function (which is typically restricted to a set of non-parametric isotonic functions, such as step-wise constants).  A drawback to both of these methods is that they assume a two-class problem; a straightforward way around this is to treat the multi-class problem as $C$ one-versus-all classification problems, where $C$ is the number of classes.  However, we find that Platt Scaling is too restrictive of a transformation to reasonably calibrate our data and determine that we do not have enough training data in each class to use Isotonic Regression with any degree of confidence.

Ultimately, we find that a calibration method similar to the one introduced by \citet{bostrom2008calibrating} is the most effective for our data.  This method uses the probability transformation
\begin{equation}
\widehat{p}_{ij} =
\left\{ \begin{array}{ll}
  p_{ij} + r (1 - p_{ij}) & \textrm{if } p_{ij} = \max\{p_{i1}, p_{i2},...,p_{iC}\} \\
  p_{ij} (1 - r) & \textrm{otherwise}
  \end{array} \right.
\end{equation}
where $\{p_{i1}, p_{i2},...,p_{iC}\}$ is the vector of class probabilities for object $i$ and $r \in [0,1]$ is a scalar.  Note that the adjusted probabilities, $\{\widehat{p}_{i1}, \widehat{p}_{i2},...,\widehat{p}_{iC}\}$, are proper probabilities in that they are each between 0 and 1 and sum to unity for each object.  The optimal value of $r$ is found by minimizing the Brier score (\citealt{brier1950verification}) between the calibrated (cross-validated) and true probabilities\footnote{The Brier score is defined as $B(\widehat{p}) = \frac{1}{N} \sum_{i=1}^N \sum_{j=1}^C (I(y_i = j) - \widehat{p}_{ij})^2$, where $N$ is the total number of objects, $C$ is the number of classes, and $I(y_i=j)$ is 1 if and only if the true class of source $i$ is $c$.}.  We find that using a fixed value for $r$ is too restrictive and, for objects with small maximal RF probability, it enforces too wide of a margin between the first- and second-largest probabilities.  Instead, we implement a procedure similar to that of \citet{bostrom2008calibrating} and parameterize $r$ with a sigmoid function based on the classifier margin, $\Delta_i = p_{i,\textrm{max}} - p_{i,\textrm{2nd}}$, for each source, 
\begin{equation}
r(\Delta_i) = \frac{1}{1 + e^{A \Delta_i + B}},
\end{equation}
where the Brier score is minimized with respect to both $A$ and $B$.  This parametrization allows the amount of calibration adjustment to differ between objects with confident (high-margin) and ambiguous (low-margin)  classifications.  Indeed, as expected, we find that the proper amount of adjustment is low for stars with small RF margin (e.g., $r(0.05) = 0.18$) and higher for sources with large maximal RF probability (e.g., $r(0.9) = 0.95$).  The parameters that minimize the Brier score over the training set are $A^* = -5.271$ and $B^* = 1.754$.

\begin{figure}
\begin{center}
\includegraphics[angle=0,width=6in]{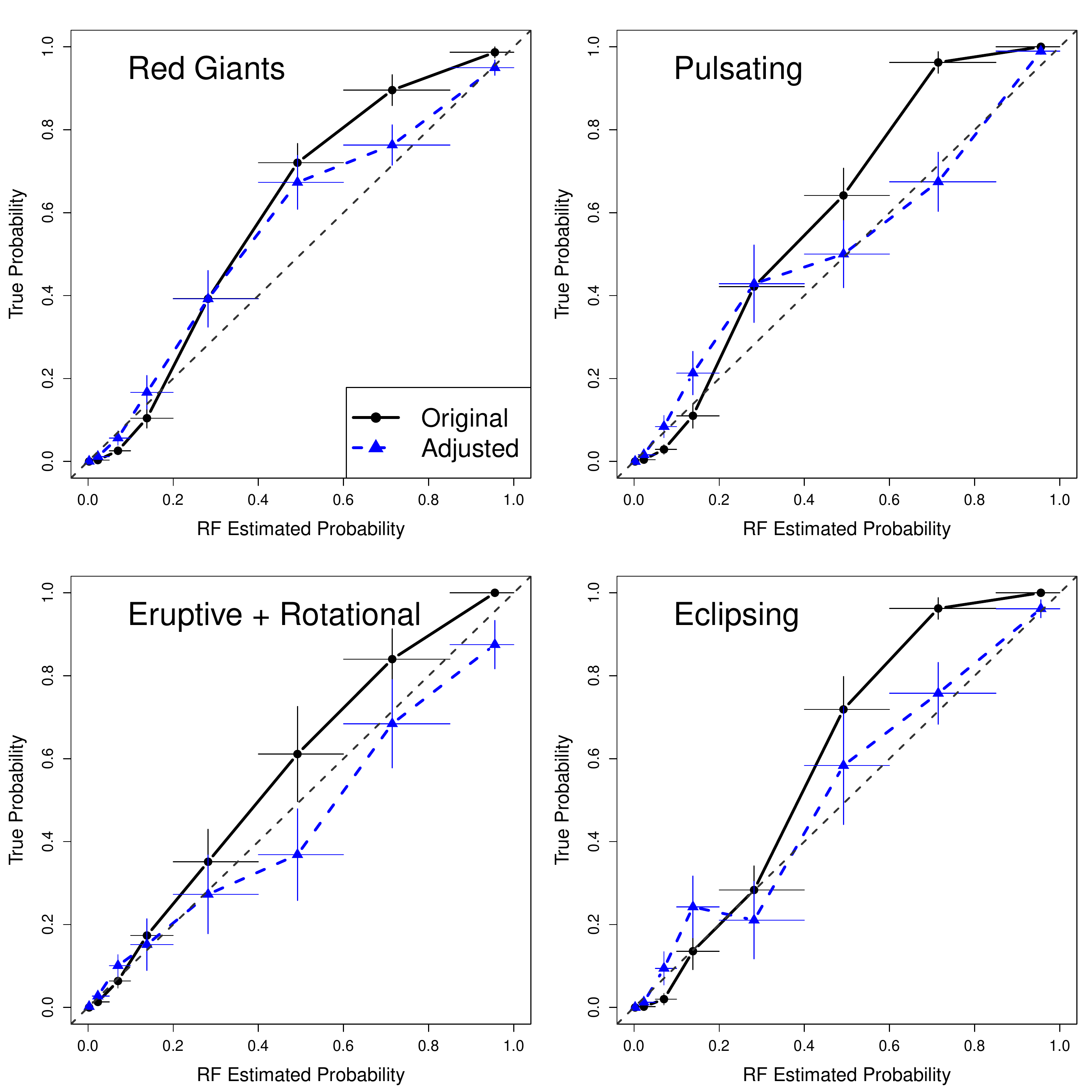}
\end{center}
\caption{ Reliability diagrams for each of four subclasses in the ASAS training data.  Within each subclass, the calibration procedure (dashed blue lines) produces  better calibration than the raw, uncalibrated random forest estimates (solid black lines).  Whereas the off-the-shelf random forest probabilities are systematically too conservative for large estimated probabilities within each of the four subclasses, the adjusted probabilities are more consistent with the diagonal for most probability bins. \label{fig:calibrationsub} }\end{figure}

 With the \citet{bostrom2008calibrating} calibration procedure, we correct the RF probability estimates for all ASAS sources.  To test the efficacy of our procedure, we plot, in the blue dashed line in Figure \ref{fig:calibration}, the adjusted (cross-validated) RF probabilities versus true posterior probabilities for our set of 777 ASAS training set objects.  The calibration is now substantially improved over the raw random forest probabilities and the calibrated probabilities are consistent with the true posterior class probabilities.  Note that the adjusted probabilities are still slightly conservative in that, on average, the estimated probabilities are systematically smaller than the true probabilities for estimated probabilities greater than ${\sim}0.1$.   In Figure \ref{fig:calibrationsub}, we plot these reliability diagrams for each of four subclasses of variable stars.  Within each of the four subclasses, the calibration has improved, with marked decrease in the Brier score for each subclass.

\subsection{Detecting Anomalous Objects}
\label{ss:anomscore}

Our calibrated ASAS probabilistic classification catalog supplies, for each object, its posterior probability of belonging to each of 28 science classes given its observed ASAS light curve and colors.  These posterior class probabilities assume prior class probabilities given by the distribution of object types in the training set (see Table \ref{tab:training}).  The posterior probabilities also assume that the training set is fully representative of the set of ASAS data, meaning that all classes present in the ASAS data are represented in the training data and that the distribution of ASAS features is the same as the training set feature distribution.  However, there is no guarantee that these conditions will be satisfied for each ASAS object, even after performing several rounds of active learning to reduce the discrepancies between the training and ASAS data sets.

The challenge, then, is to identify ASAS objects which do not resemble any of the training data.  Classifier predictions for these objects will be dubious due to the outlying nature of their feature vectors compared to the training set feature distribution, either due to their belonging to a class not included in the training set or anomalous features brought about by noise or atypical physical variability.  To detect such anomalies, we compute, for each ASAS object, a distance metric from that object's feature vector to each source in the training set.  In contrast to previous methods which compute distances between phased light curves for periodic variable stars to detect anomalies (\citealt{2006MNRAS.369..677P,2009ASPC..411..264R}), we compute a distance measure between feature vectors.

Similar to \citet{2012sharmo}, we use a semi-supervised approach to compute the anomaly score for each variable star.  We begin by fitting a random forest classifier to the training set as in \S \ref{ss:train}.  The random forest outputs a proximity measure $\rho_{ij}$, between each pair of sources $i$ and $j$, which gives the proportion of trees in the random forest for which the feature vectors $\x_i$ and $\x_j$ appear in the same terminal node.  If two sources have similar feature vectors with respect to the topology of the random forest, then the proximity will be near 1, whereas if the feature vectors are dissimilar then the proximity will be near 0.  Using the proximity measure, we define the discrepancy between the two feature vectors $\x_i$ and $\x_j$ as 
\begin{equation}
\label{eqn:anom}
d(\x_i,\x_j) = \frac{1 - \rho_{ij}}{\rho_{ij}}
\end{equation}
which takes on non-negative real valued numbers.  This metric is semi-supervised because it uses the labeled training set to construct the optimal random forest classifier, which is then used to compute proximities (and discrepancies) between labeled and unlabeled sources.  

The novelty of the distance measure in Equation \ref{eqn:anom} is that it automatically gives more weight to features which are important in the classifier while ignoring useless features.  For instance, if a feature is important for classification, then the RF trees will make many splits on that feature, thus dividing the coordinate into many sub-regions.  Hence, for a new source, the value of that class-predictive feature will have a great deal of power in determining which terminal node the source falls into for each tree, and thus will be a strong determinant of its proximity to other sources.  Likewise, features that are unimportant for classification will never be split on by any tree, and thus proximities will be unaffected by their values.  Unlike Euclidean distance, the proximity-based distance measure adapts to the geometry of the classification problem and can treat different regions of feature space differently based on the class boundaries and prevalence of training data in those regions.

Using the RF proximity measure, we  construct an anomaly score for each ASAS object.  We first compute the distance, using Equation \ref{eqn:anom}, from the feature vector of each ASAS source to the feature vector of every training source.  We define the anomaly score for each ASAS object to be the distance (Equation \ref{eqn:anom}) to the 2nd nearest neighbor in the training set.  Objects with large anomaly scores should be considered as outliers and their classifications should not be trusted because there is too much discrepancy between the features of those  sources and the set of training set variable stars.  Note the subtle difference between the anomaly score and classification probability: sources with small maximal class probability may reside near training data but fall in regions of feature space shared by several science classes.  At the same time, sources with high anomaly score may have large maximal class probability due to their relatively close proximity to the training objects of a certain class compared to the training objects of the other classes.

\begin{figure}
\begin{center}
\includegraphics[angle=0,width=6in]{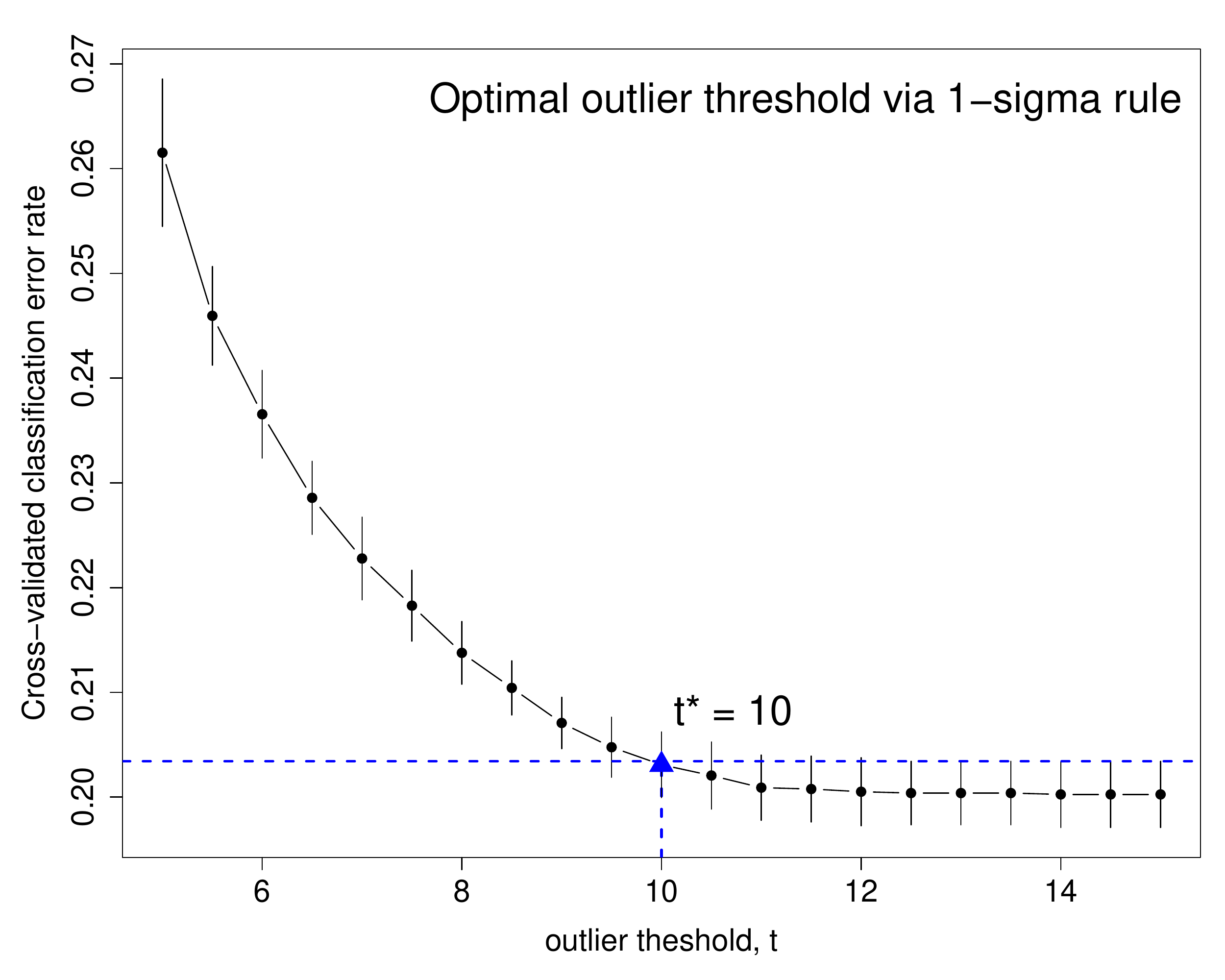}
\end{center}
\caption{ Determination of the optimal anomaly score threshold via cross-validation.  As the outlier threshold, $t$, is reduced, more objects are considered anomalies, and the cross-validated error rate increases (outliers are, by construction, assigned no label, incurring a classification penalty of 1).  Using the 1-$\sigma$ rule, which chooses the smallest threshold for which the error rate is within one standard deviation of the default model with no thresholding, we find that the optimal threshold on anomaly score is $t^* = 10.0$.  Adopting this threshold for the ASAS data, we discover 1741 outliers.   \label{fig:anomscore} }\end{figure}

The anomaly score provides a positive real-valued number for each object.  However, we may ultimately want to make a decision, for each object, of whether or not that source is an outlier, by thresholding on the anomaly score.  To determine an appropriate score threshold for anomaly detection, we employ cross-validation on the training set.  In each of $K=10$ cross-validation folds, we hold out a random subset of the ASAS training data, fit the random forest classifier on the remaining data, and compute the anomaly score for each held-out object.  Then, for each anomaly score threshold, we record the cross-validated classification error rate over the ASAS training data, counting each object whose anomaly score surpasses the threshold as an error.  Results of this experiment are in Figure \ref{fig:anomscore}.  As the threshold decreases, we identify more objects as outliers, but the classification error rate only becomes significantly affected for thresholds smaller than 10.0.  Following the 1-$\sigma$ rule of \citet{2009hast} over 10 repetitions of the procedure, we find that the optimal threshold level is $t^* = 10.0$.  Therefore, we recommend that the 1741 ASAS objects with anomaly score larger than 10.0 be treated as outliers.

\begin{figure}
\begin{center}
\includegraphics[angle=0,width=6in]{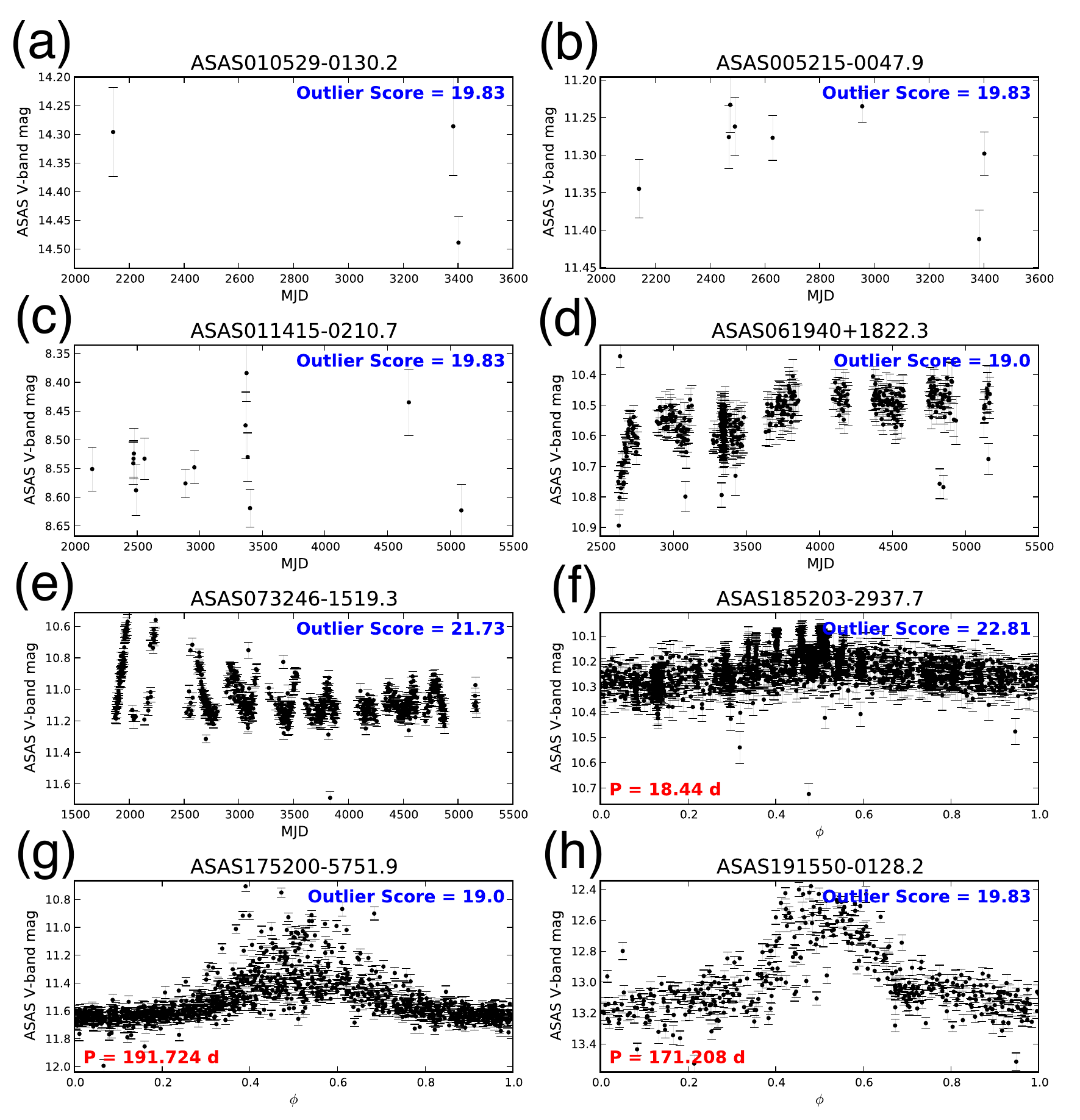}
\end{center}
\caption{ Top outlier light curves as determined from the anomaly score.  These light curves are the furthest from their second nearest neighbor in the training set according to the anomaly metric in Equation \ref{eqn:anom}.  These sources are outliers due to either too few data (a,b,c), long-scale secular trends (d), quasi-periodic behavior of irregular type (e), occupying an anomalous region of period--amplitude space (f), or high-amplitude variability suppressed due to blending with nearby sources (g,h).  For each of these sources, there are no training instances that capture the observed variability in their ASAS light curves.     \label{fig:outliers} }\end{figure}

In Figure \ref{fig:outliers} we plot the ASAS light curves of eight sources that are amongst the highest outlier scores.  These objects include a light curve with only 3 epochs of data, two other light curves with with fewer than 15 epochs of data, a known emission-line star showing variability on long time scales (ASAS061940+1822.3), a likely Be star showing semi-regular pulsations with amplitude modulation (ASAS073246-1519.3), and a red star with quasi-periodic low-amplitude variability on 18-day timescales (ASAS185203-2937.7).  Additionally, two of these objects show semi-regular pulsations of smaller than 1 mag on 150--200-day timescales (ASAS175200-5751.9 and ASAS191550-0128.2); these are likely to be blended Mira variables.  For each of these outliers, there are no training instances that capture the observed variability in their ASAS light curves.

\section{The Catalog}
\label{sec:catalog}

Here we describe the contents of the publicly-available Machine-learned ASAS Classification Catalog.  MACC is available for download at \url{www.bigmacc.info}.   The first 40 rows of the classification catalog are reproduced here in Table \ref{tab:catalog}.  The columns of the catalog are as follows:
\begin{itemize}
\item \verb ASAS_ID  -  ID from ASAS catalog of Variable Stars
\item \verb dotAstro_ID  -  ID from the online database \url{http://dotastro.org/}
\item \verb RA , \verb DEC  -  Coordinates from ASAS\footnote{Coordinates from ASAS are sometimes wrong by several arcsec due to its $\sim$15-arcsec pixel size.  This effect is worse in crowded fields.}
\item \verb Class  -  Most probable class from the machine learned classifier
\item \verb P_Class  -  Posterior probability that the source is from that class, given the ASAS light curve and colors
\item \verb Anomaly  -  Metric from \S\ref{ss:anomscore}; objects with score greater than 10.0 should be considered as outliers
\item \verb ACVS_Class  -  Classification from the ASAS Catalog of Variable Stars (\citealt{2002AcA....52..397P})
\item \verb Train_Class  -  If the ASAS object was in the training set, its training class, else blank
\item \verb Mira ,..., \verb W_Ursae_Maj  -  Posterior class probabilities for all 28 science classes
\item \verb P , \verb P_signif  -  Best fit period (in days) and its statistical significance (in number of $\sigma$)
\item \verb N_epochs  -  Number of epochs in the ASAS light curve used to classify the object
\item \verb V , \verb deltaV  -  Mean ASAS $V$-band magnitude and ASAS $V$-band amplitude
\end{itemize}

MACC has been constructed to allow easy querying of objects of a specified science class, simple searching for outliers, and more advance queries on several attributes.  In supplying the posterior class probabilities for each class,  the catalog allows each individual researcher to define their own probability threshold in querying objects.  For instance, imagine that scientists A and B are both interested in finding Mira variables, but scientist A requires a highly pure sample while scientist B simply wants the top 3,000 Mira candidates, even if a substantial number of these are non-Miras.  Then, scientist A could use a strict threshold, selecting all candidates with $\Prob$(Mira) $>0.95$ and Anomaly $<10$ (resulting in 2,131 very likely Mira candidates), while scientist B would simply grab the 3,000 objects with largest $\Prob$(Mira) (which, in this case is equivalent to a Mira probability threshold of 0.470).

\begin{deluxetable}{lllllllllllllllll} 
\tablecolumns{17}
\tabletypesize{\tiny}
\rotate 
\tablewidth{58pc} 
\tablecaption{First 40 rows of the Machine-learned ASAS Classification Catalog. The entire table is available at \url{www.bigmacc.info}.}
\tablehead{ 
\colhead{ASAS ID } & \colhead{dotastro\tablenotemark{a}} & \colhead{RA\tablenotemark{b}} & \colhead{DEC\tablenotemark{a}} & \colhead{Class} & \colhead{P(Class)} & \colhead{Anomaly} & \colhead{ACVS Class} & \colhead{Train Class} &\colhead{P(Mira)} & \colhead{...}  & \colhead{P} & \colhead{P signif\tablenotemark{c}} & \colhead{N } & \colhead{V\tablenotemark{d}} & \colhead{$\Delta$V\tablenotemark{e}} }
\startdata 
\input{asas_catalog.dat}
\enddata
\tablenotetext{a}{ID from the online database \url{http://dotastro.org/}}
\tablenotetext{b}{In decimal degrees.}
\tablenotetext{c}{Statistical significance of the period against a null hypothesis of white noise, in number of $\sigma$.}
\tablenotetext{d}{Average V-band magnitude.}
\tablenotetext{e}{Peak-to-peak amplitude (95th minus 5th quantile).}
\label{tab:catalog}
\end{deluxetable}%

\begin{figure}
\begin{center}
\includegraphics[scale=0.4]{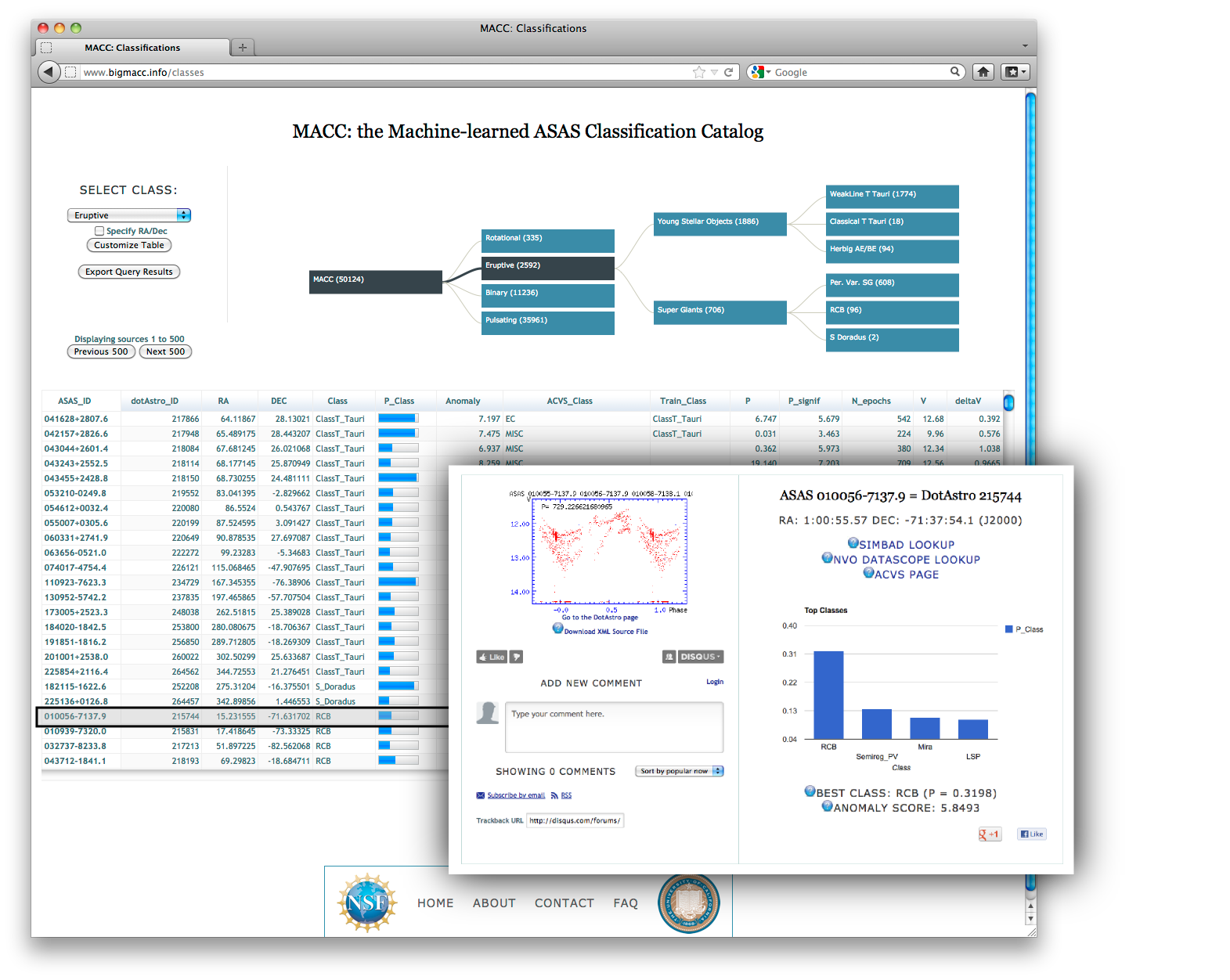}
\end{center}
\caption{ A screenshot of the \textbf{BROWSE} section of the \url{http://www.bigmacc.info} website, the online version of the Machine-learned ASAS Classification Catalog. Here, we show a subsection of the classification taxonomy  where the user has selected the \emph{Eruptive} branch, which is also reflected in the table below and the drop-down in the top-left corner. The embedded figure shows the information page for an individual source, which provides light curves, class probabilities and comments, along with other meta-data. \label{fig:website} }\end{figure}

In addition to a full catalog download, the online catalog at \url{www.bigmacc.info} provides several ways for users to interact with the data. For users familiar with the SQL query language, the \textbf{QUERY} page enables users to specify SQL queries with conditions of their choice. A \emph{Basic query} provides the simplest SQL interface for the data, while the \emph{Advanced query} allows fine-grained control of the output.
The \textbf{BROWSE} page allows users to explore the catalog by the most probable class determined by our classifier. A visualization of the tree of science classes enables a novel way of exploring the variable star taxonomy while querying and presenting the data in real-time.
Examples of visualization of the data are given in the \textbf{VISUALIZE} section along with guides on how to create these and other plots without downloading the data.  See Figure \ref{fig:website} for a screenshot of the website.

In all the tables on \url{www.bigmacc.info}, clicking a table header will sort the data by that column and customization of the columns can be done in the \emph{Customize Table} dialog. Certain circumstances might require astronomers to query only a certain area on the sky, which is supported by the \emph{Specify RA/Dec} dialog where the RA, Dec and a search radius can be specified. Clicking single rows will show the individual source page that provides important information about the source including light curves, most probable member classes and a comments section to encourage feedback and collaboration within the variable star research community. 
For offline access to query results, the \emph{Export Query Results} button will give the user a traditional CSV file for download. 

The online catalog has been built on top of the \href{http://www.google.com/fusionHome/}{Google Fusion Tables} framework, which allows quick and robust access to the data while minimizing administration overhead associated with hosting a catalog like this. The sky position of each source is geocoded and indexed in the master fusion table to enable fast position-constrained queries.  The front page provides full access to the \emph{Fusion Tables} interface where queries and visualizations can be customized to the users needs. Furthermore, the site is built on the \href{https://developers.google.com/appengine/}{Google App Engine} for \href{http://www.python.org/}{Python}, a cloud-based solution providing free hosting of web applications below a certain usage limit and the ability to scale on demand in case of temporary or permanently increased traffic. These services, along with powerful client-side Javascript frameworks such as \href{http://jquery.com/}{jQuery} and \href{http://thejit.org/}{InfoVis Toolkit} allow rapid development of data-centric web application with minimal administrational constraints, which are all important aspects in the presentation of a catalogue like this.

\subsection{Substituting Different Class Priors}
\label{ss:priors}

All of the posterior class probabilities given in MACC assume that the prior probability of observing an object of class $c_j$ (before observing any data) is given by the proportion of training set objects that are of class $c_j$ (provided in Table \ref{tab:training}).  By Bayes' Rule, the posterior MACC class probability for class $c_j$ given the features, $\x_i$, for object $i$, is
\begin{equation}
\Prob(c_j | \x_i) = \frac{\Prob(\x_i | c_j) \Prob_{\mathrm{tr}}(c_j)}{ \sum_{k=1}^{28} \Prob(\x_i | c_k)  \Prob_{\mathrm{tr}}(c_k)}
\label{eqn:bayes}
\end{equation}
where $\Prob_{\mathrm{tr}}(c_j)$ is the prior class probability given by the proportion of objects of class $c_j$ in the training set.  To exchange a different vector of prior class probabilities, one must multiply each posterior probability from the catalog by the ratio of the new prior to the training set prior  and multiply by the corresponding ratio of denominators from Equation \ref{eqn:bayes}.  For a new prior $\Prob_{\mathrm{new}}(c_j)$, the new posterior probabilities are given by
\begin{equation}
\label{eqn:newprior}
\Prob_{\mathrm{new}}(c_j | \x_i) = \Prob(c_j | \x_i) \frac{\Prob_{\mathrm{new}}(c_j)}{\Prob_{\mathrm{tr}}(c_j)} \frac{\sum_{k=1}^{28} \Prob(\x_i | c_k)  \Prob_{\mathrm{tr}}(c_k)}{\sum_{k=1}^{28} \Prob(\x_i | c_k)  \Prob_{\mathrm{new}}(c_k)}.
\end{equation}
For modified priors, $\Prob_{\mathrm{new}}$, that are not too dissimilar from $\Prob_{\mathrm{tr}}$, the last term in Equation \ref{eqn:newprior} will typically be near unity, and thus the modified posterior probabilities can be computed by multiplying the original posteriors by the prior ratio and appropriately re-normalizing.  For very dissimilar priors, accurate estimates of all of the class-wise densities, $\Prob(\x_i | c_k)$, would have to be computed and stored on a fine grid of the 71-dimensional feature space, which is both statistically and  computationally infeasible \footnote{Consider the most na\"{i}ve density estimate, a histogram.  Constructing a 71-dimensional histogram for each class by binning each feature into 10 bins requires $28 \times 10^{71}$ values to be computed and stored.  Statistically, such a density estimate is unreliable, as the amount of training data is microscopic compared to the vast feature space occupied by 71 dimensions, rendering any simple density estimate useless.}. In the absence of satisfactory estimates of  the class-wise densities, and short of re-training the random forest classifier with different prior weights, it is reasonable to update the posteriors by assuming that the last term in Equation \ref{eqn:newprior} is unity.

The construction in Equation \ref{eqn:newprior} allows us to also condition on additional information such as galactic coordinates $(\ell, b)$, median magnitude, and/or distance.  For instance, if we have a good theoretical understanding of the expected demographics of variable stars as a function of position in the galaxy, we can imbue that information into the prior probabilities.  In other words, before observing any data for a particular object, we can modify its prior class probabilities solely based on its location in the galaxy.  This can be a very powerful tool, e.g., for finding star-forming regions near the galactic plane, where the relative abundance of young stellar objects will be higher (and that information can be inserted into the class prior).

\subsection{Difficult Class Boundaries}

\begin{figure}
\begin{center}
\includegraphics[angle=0,width=6in]{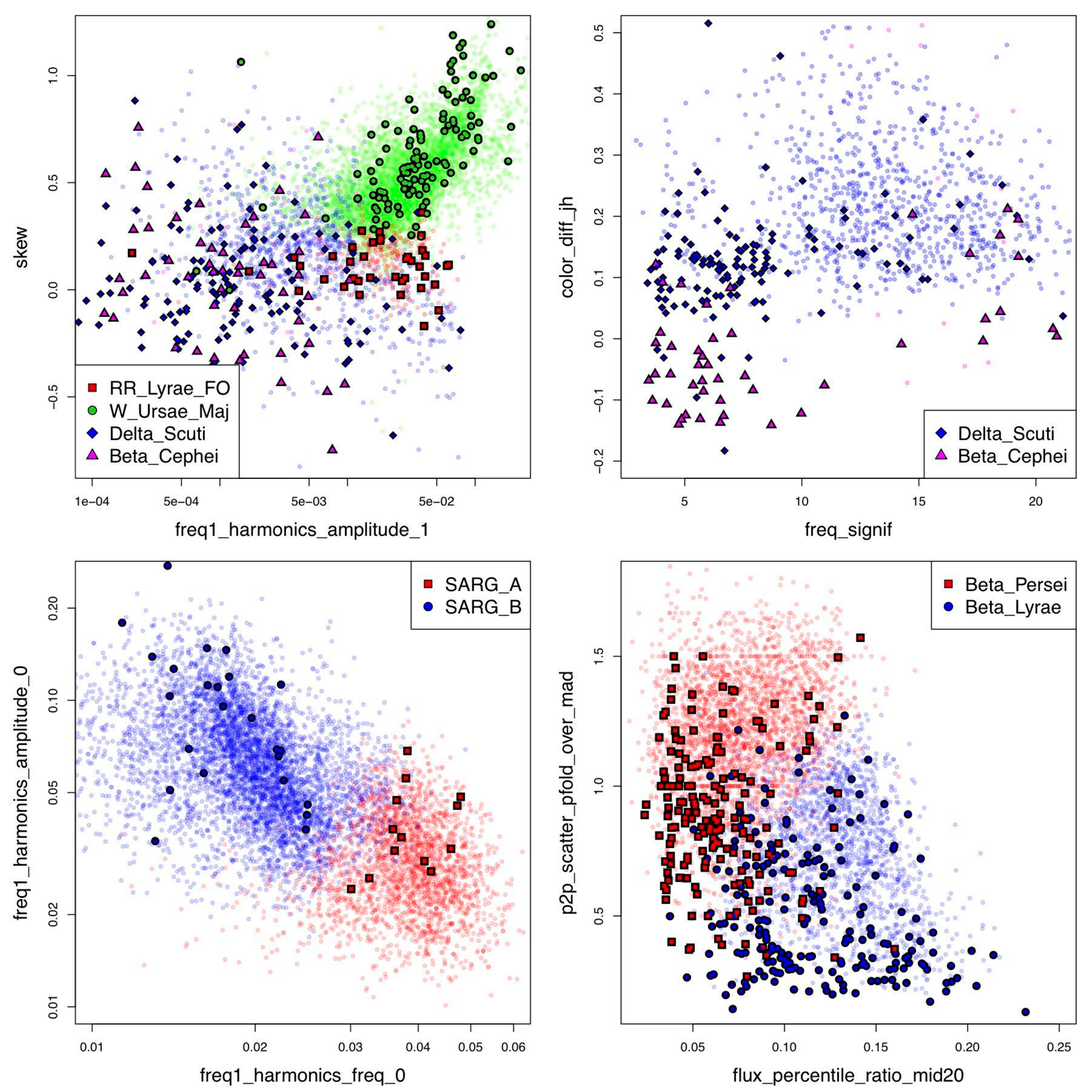}
\end{center}
\caption{ The Random forest classifier automatically discovers class boundaries in the high-dimensional feature space.  For certain easily confused classes, we plot the projections, in two-dimensional feature spaces, of training objects (points with solid outline) and  MACC-classified objects (small dots).   Top Left: In the skew--first-harmonic-amplitude  plane, W Ursae Majoris, RR Lyrae FO, and Delta Scuti stars are well separated, but Delta Scuti and Beta Cephei remain confused.  Top Right: However, Delta Scuti and Beta Cephei are separated by their $J-H$ color.  Bottom Left: SARG A and B subtypes split naturally in the period-amplitude plane.  Bottom Left: Beta Persei and Beta Lyrae binaries are largely separable by two features, with a small amount of overlap. \label{fig:boundaries} }\end{figure}

There are certain classes of variability that are difficult to separate based on photometric information alone.  For instance, W Uma Majoris, Delta Scuti, and RR Lyrae, FO stars all show variability on the same time scales with similar amplitudes.  Other classes, such as Weak-line T Tauri and RS CVn stars exhibit variability from similar physical mechanisms (in this case, rotation of chromospherically active stars), which may result in ambiguous classification of sources of those classes based on light curve information alone.  An advantage of using ML classification is that, given enough training data, these methods can learn which light curve features best separate sources of similar class and can determine optimal class boundaries.  In Figure \ref{fig:boundaries}, we plot the most informative features for separating notoriously difficult-to-separate classes of variable star.  Even with relatively few training instances, the classifier effectively learns how to best distinguish, e.g., Delta Scuti and Beta Cephei.

\begin{figure}
\begin{center}
\includegraphics[angle=0,width=6in]{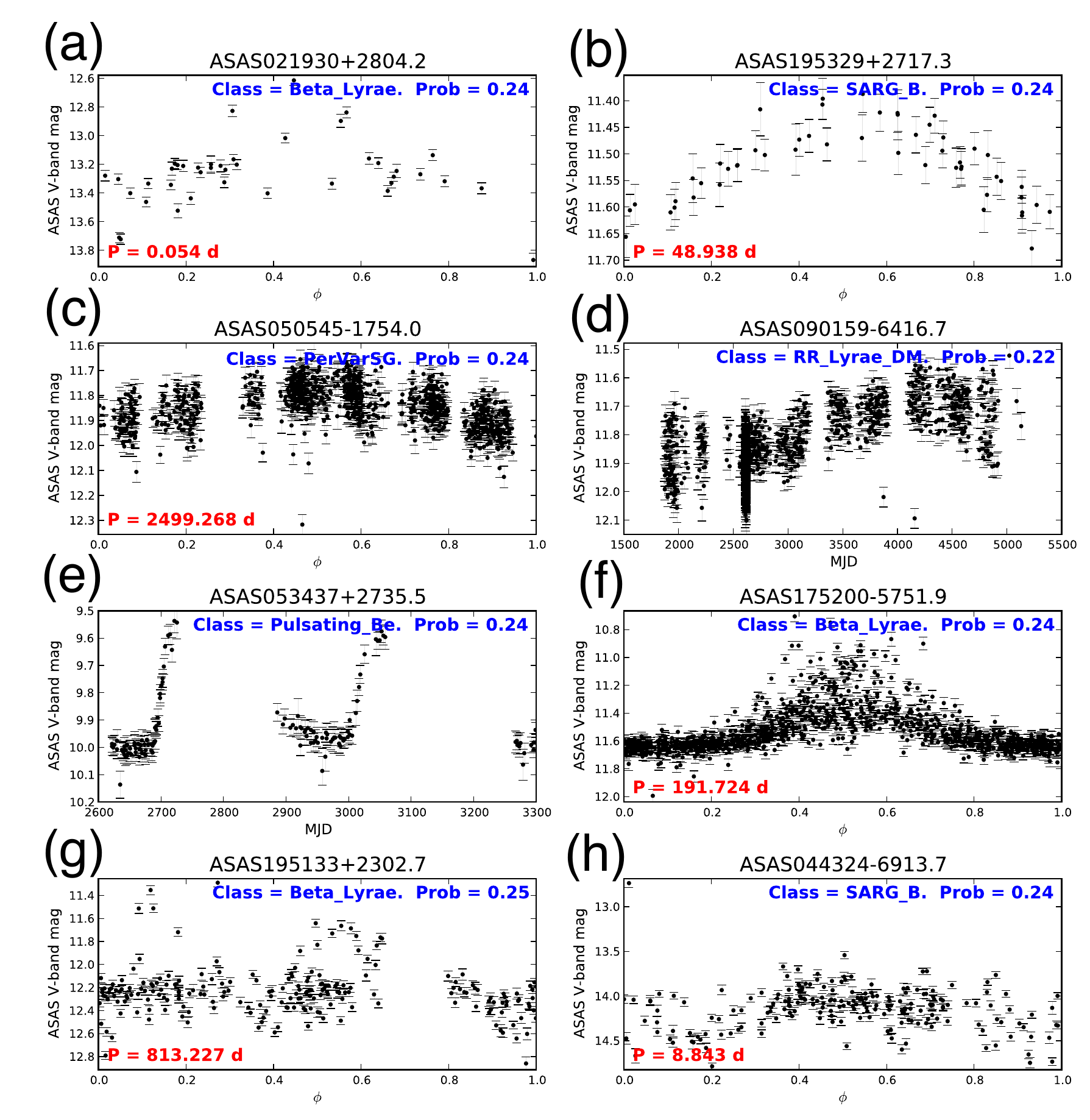} 
\end{center}
\caption{ ASAS light curves for the candidates with lowest classification probability across all 28 classes.  Several of these light curves suffer from lack of data (a), large temporal gaps (e), or large amounts of noise caused either by blending with nearby stars (f,g) or relative faintness (h).    Others are outliers due to abnormal period-amplitude combinations (b), or secular variability on several year timescales (c,d).
These objects---and others that obtain low probabilities across all 28 science classes---require further study to ascertain their true nature.   \label{fig:noconf} }\end{figure}

\begin{figure}
\begin{center}
\includegraphics[angle=0,width=6in]{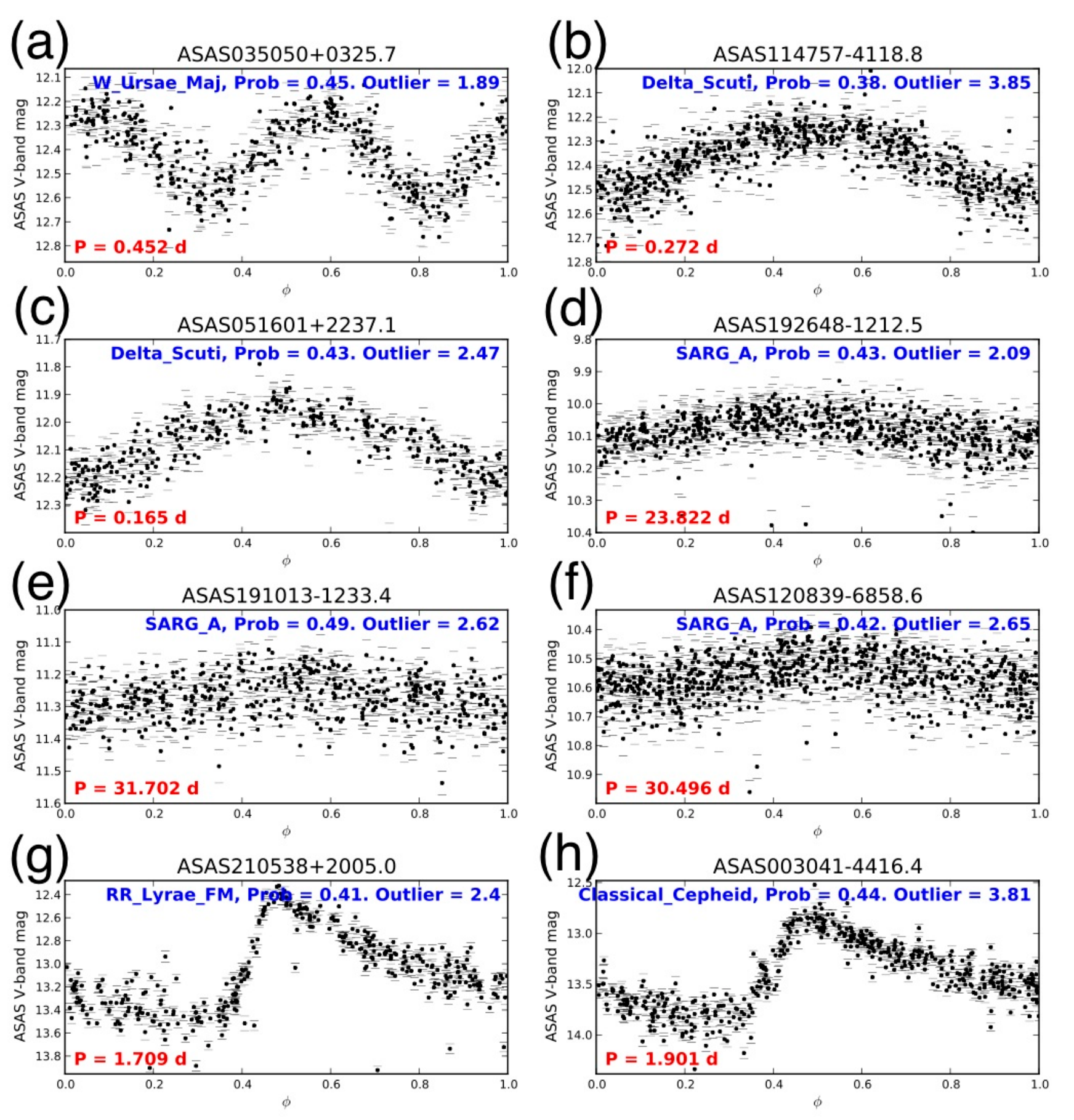}
\end{center}
\caption{ Light curves of ASAS objects whose anomaly scores are small even though their maximal classification probabilities are smaller than 0.5.  These light curves show behavior not inconsistent with a particular class of variability but typically reside between classes.  The objects in (a,b,c) reside on the border between Delta Scuti, W Uma Majoris, and RR Lyrae, FO variable stars.   Likewise, the stars in (d,f) lie on the boundary between SARG A and B subtypes.  Similarly, the objects in (g,h) are either Classical or Population II Cepheids.   \label{fig:notoutliers} }\end{figure}

That said, there will always be borderline cases, for which, given their light curve and color data, it is impossible to confidently place the objects into a class.  This uncertainty is reflected  by low posterior class probabilities assigned by the classifier across all classes.  In Figure \ref{fig:noconf}, we plot the ASAS light curves for a few of the least confidently classified (lowest maximal posterior probability) sources in MACC.  These sources typically have poor data quality and/or fall in outlying regions of light-curve feature space, meaning that there is not enough light curve information from these objects for the classifier to make a confident statement about their class.
For comparison, in Figure \ref{fig:notoutliers} we plot a few of the ASAS objects whose light curves have low outlier score but whose highest posterior class probability is smaller than 0.5.  These light curves do not show atypical behavior, but tend to reside on the boundary between classes.  The objects ASAS 035050+0325.7, ASAS 114757-4118.8, and ASAS 051601+2237.1 all reside on the W Uma--Delta Scuti locus, making them impossible to classify with any degree of confidence.  Likewise, ASAS 192648-1212.5, ASAS 191013-1233.4, and ASAS 120839-6858.6 all reside near the boundary between the SARG A--B classes.
Also, both ASAS 003041-4416.4 and ASAS 210538+2005.0 are Cepheids with atypically high amplitudes and short periods that place them near the dividing line between Classical and Population II Cepheid stars.

\section{Comparison to Literature}
\label{sec:compare}

We conclude with a comparison of our Machine-learned ASAS Classification Catalog with a set of papers that have performed classification for ASAS objects.  We first analyze the similarities and differences between our classification catalog and the popular ACVS catalog.  Subsequently, we take a closer look at a handful of papers that have attempted to find, in the ASAS data, objects of specific sub-classes.  Overall, we find a very high classification agreement rate with these other works.  For cases in which our classification catalog disagrees with the class-specific papers, the differences can be attributed to poor quality of the ASAS photometry and extra information that was unavailable to our classifier such as proprietary follow-up data including spectra and high-quality multi-band light curves.

\subsection{ASAS Catalog of Variable Stars (ACVS)}

As a part of the ASAS Catalog of Variable Stars, predicted classes are provided for a fraction of the stars.  As described in \citet{2002AcA....52..397P}, ACVS obtains their classifications using a neural net type algorithm trained on set of  visually labeled ASAS sources, confirmed OGLE cepheids (\citealt{1999AcA....49..223U,1999AcA....49..437U}), and OGLE Bulge variable stars (\citealt{2002AcA....52..129W}).   A filter is used to divide strictly periodic from less regular periodic sources.  A neural net is trained on the period, amplitude, Fourier coefficients (first 4 harmonics), $J-H$ and $H-K_s$ colors and IR fluxes to predict the  classes of the strictly periodic sources.  Many ACVS objects either have multiple labels or are annotated as having low confidence classifications, but no posterior class probabilities are given in the catalog.  For less regular periodic sources, location in the $J-H$ versus $H-K_s$ plane is tested; if the object falls within an area of late-type irregular or semi-regular stars, it is assigned the label MISC, else it is inspected by eye.  We find that 38,117 ACVS stars, representing 76\% of the catalog,  are either labeled as MISC, assigned multiple labels, or have low class confidence.  The remaining 24\% of stars have confident ACVS labels, and provide a set of classifications to compare against our catalog.  

\begin{figure}
\begin{center}
\includegraphics[angle=0,width=5.5in]{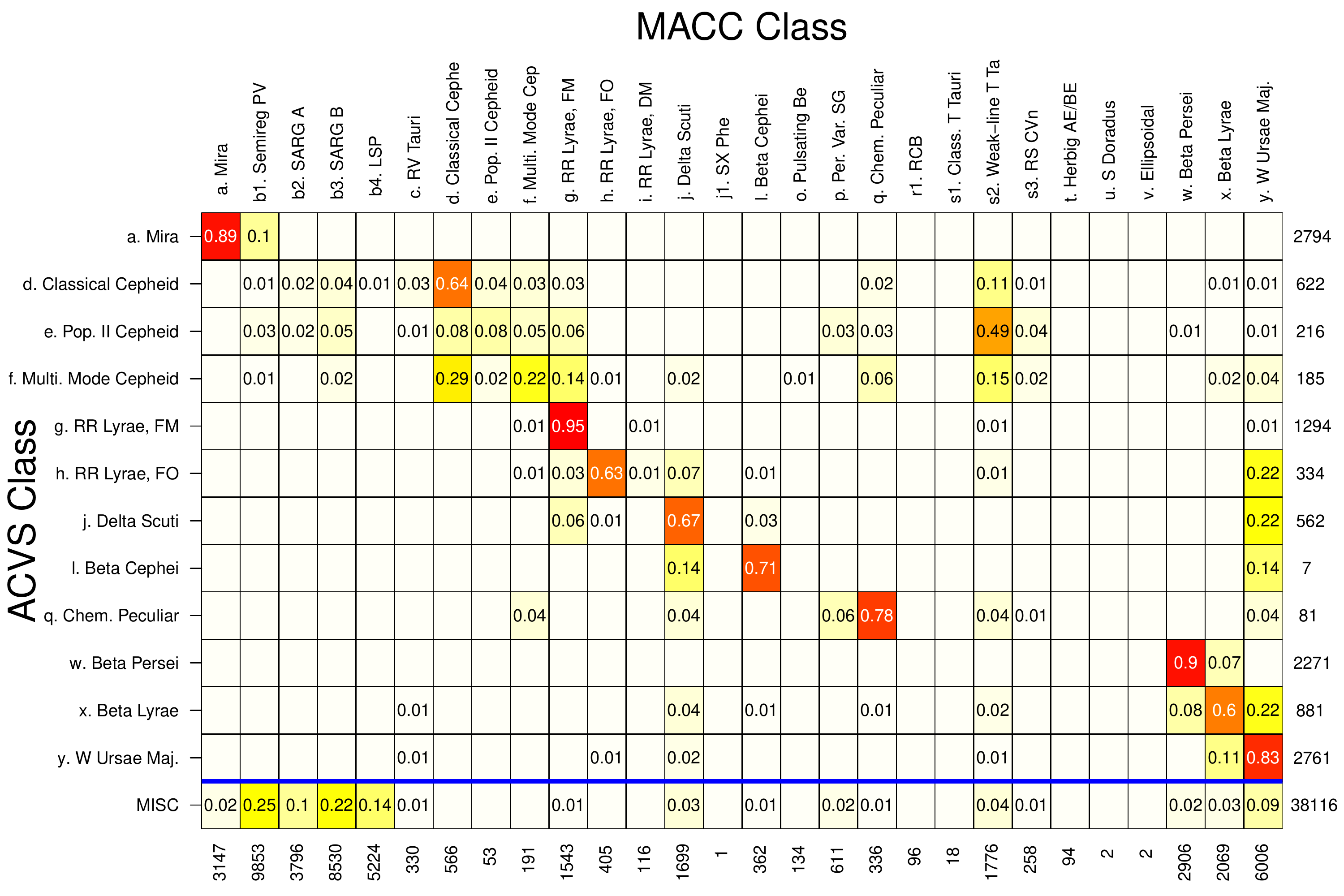}\\
\includegraphics[angle=0,width=5.5in]{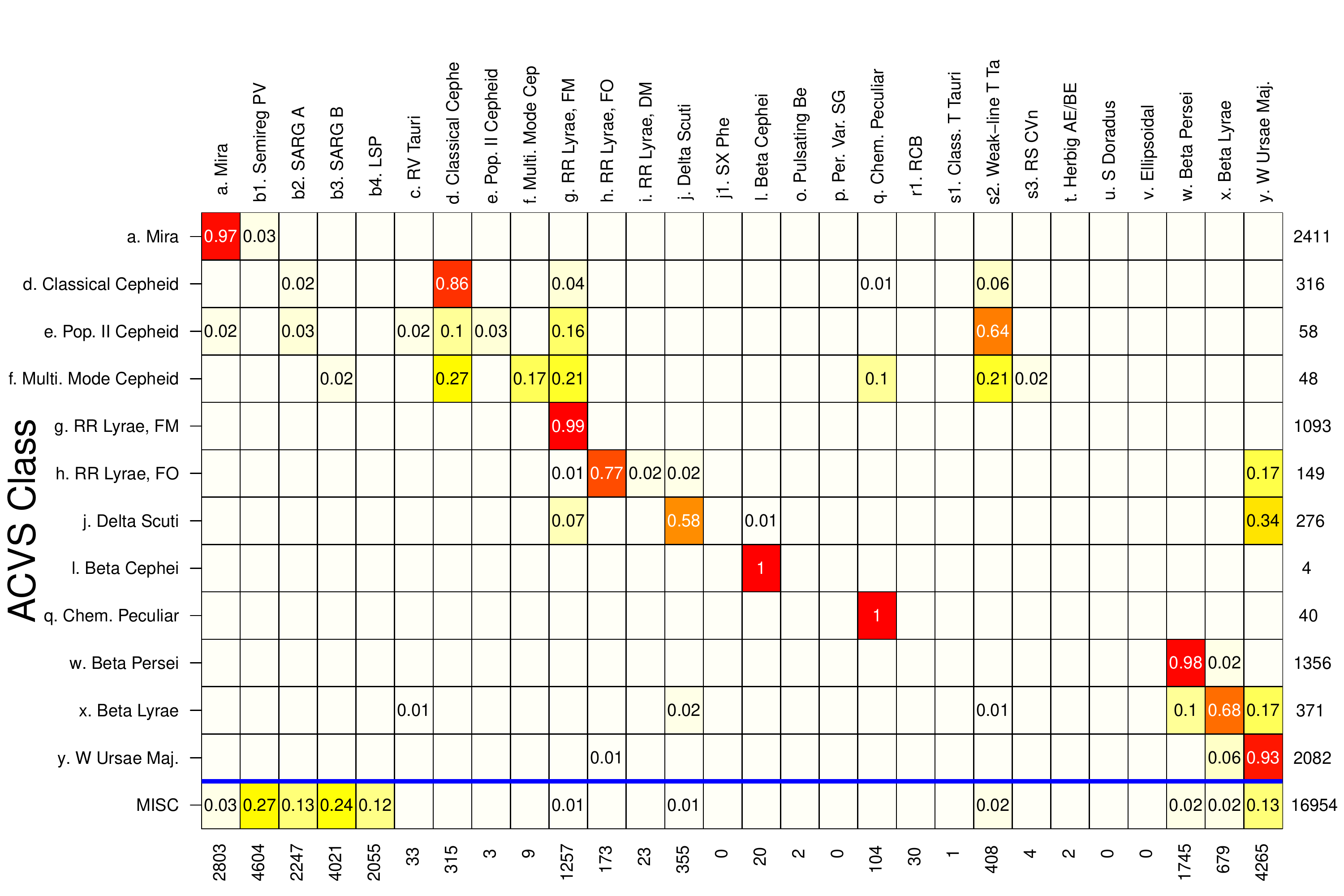}
\end{center}
\caption{Top: Correspondence of the MACC to the ACVS classifications for all 50,124 sources.    Rows are normalized to sum to 100\%, marginal counts are listed to the right and bottom of the table.  There is an 80.5\% total correspondence between our classifications and the ACVS labels, for the 12,007 objects whose ACVS classification is a single confident class not equal to MISC.
Bottom: Same for the subset of 25,158 ASAS sources with outlier score smaller than 3.  The agreement rate between MACC and ACVS for the subset of these sources with confident ACVS class (8,204 objects) is 91.6\%.
  \label{fig:acvs}}
\end{figure}

In the top panel of Figure \ref{fig:acvs} we plot the class-wise correspondence between our classifications and the ACVS classes.  Overall, there is an  80.5\% correspondence between our catalog and ACVS, for the 12,007 sources that are labeled confidently (and not as MISC) in ACVS.  For each of the ACVS sub-classes, except Population II Cepheid and Multi-Mode Cepheid, we agree on at least 60\% of objects.  The large disagreement with the Population II and Multi-Mode Cepheids is consistent with the results of \citet{2009AJ....137.4598S} who find extreme biases in Cepheid classifications for ACVS.  Of 282 stars labeled as Cepheid by ACVS, only 14 were found by \citet{2009AJ....137.4598S} to be likely Pop. II cepheids, while all but $\sim60$  suffered from uncertain period estimates, and $\sim50$ were rejected as obvious non-Cepheids.  We also find that our classifications of First Overtone RR Lyrae, Delta Scuti, and W Ursae Majoris  show a significant amount of discrepancy with those of ACVS.  In particular, our classifier finds that $\sim$22\% of the stars that ACVS classifies as RRc or Delta Scuti are truly WUma eclipsing variables.   
 
 In the bottom panel of Figure \ref{fig:acvs}, we plot the class-wise correspondence for all 25,158 ASAS sources with MACC outlier score smaller than 3.0.  For these more confidently classified objects, MACC has a closer correspondence with ACVS (91.6\% for the 8,204 objects with confident ACVS class), but still shows high level of disagreement for the non-Classical Cepheids.  Of these sources, we find a 97\% agreement on Miras, 86\% on Classical Cepheids, 99\% on RR Lyrae, FM, perfect agreement on 39 Chemically Peculiar stars, 98\% on Beta Persei, and 93\% on W Uma Majoris sources.

\subsubsection{Confident MACC Classifications missed by  ACVS}

In addition to having $>80\%$ correspondence with ACVS for objects which they confidently label, our MACC catalog identifies many confidently classified sources---having posterior class probability of at least 0.9 for a single class---whose ACVS classification is either uncertain (denoted by a `:' in the catalog) or split between multiple classes.  In all, MACC identifies 187 Mira, 22 Classical Cepheid, 122 Fundamental Mode RR Lyrae, 11 First Overtone RR Lyrae, 14 Beta Cephei, 43 Chemically Peculiar, 152 Beta Persei, 210 Beta Lyrae, and 1548 W Uma Majoris candidates that were not found by ACVS.  Lowering the confidence threshold from 0.9 to 0.8 yields about 50\% more good candidates.

\begin{figure}
\begin{center}
$\begin{array}{c@{\hspace{-.3in}}c}
\multicolumn{1}{l}{\mbox{\bf (a)}} & \multicolumn{1}{l}{\mbox{\bf (b)}} \\ 	\vspace{-.1in}
\includegraphics[angle=0,width=3.5in]{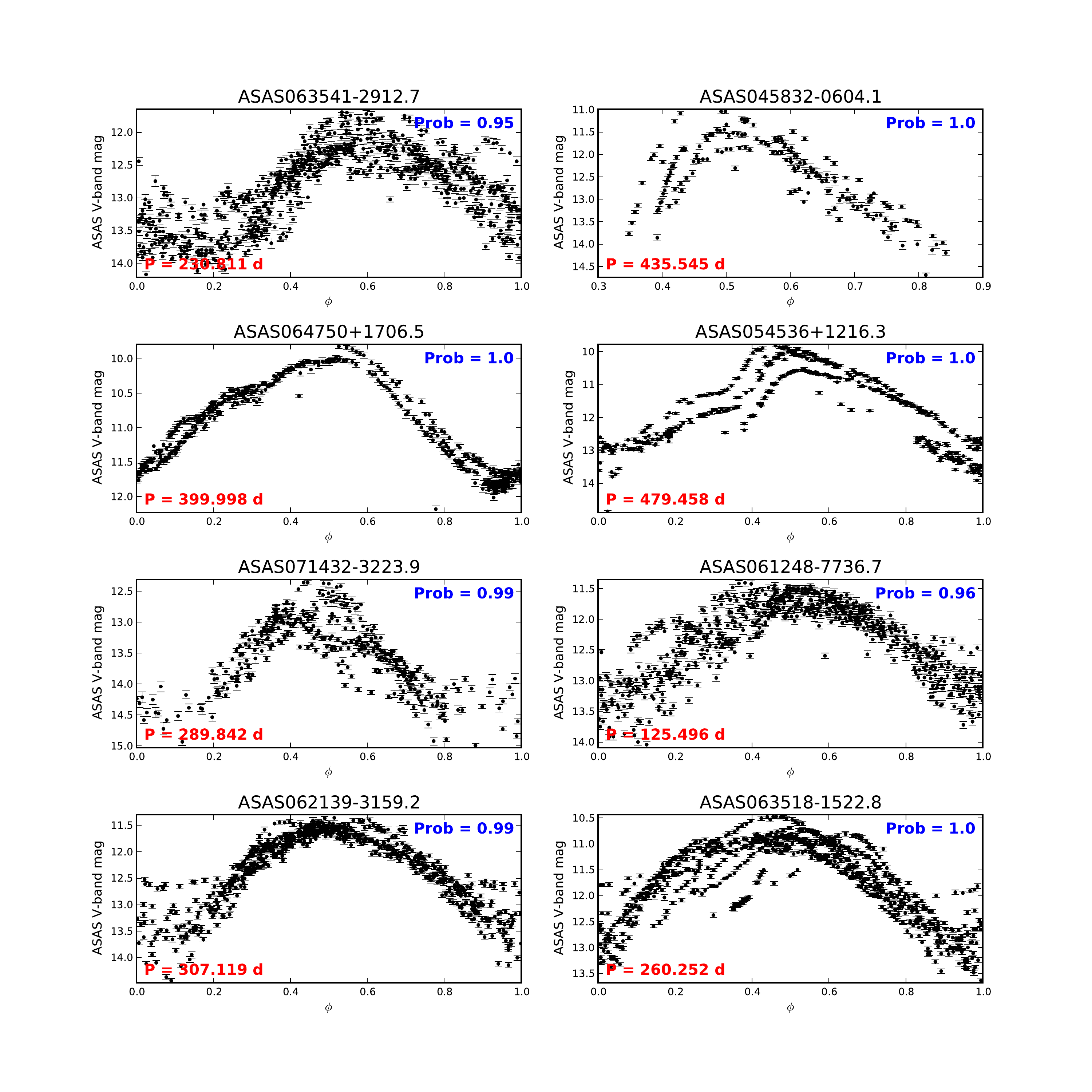} & \includegraphics[angle=0,width=3.5in]{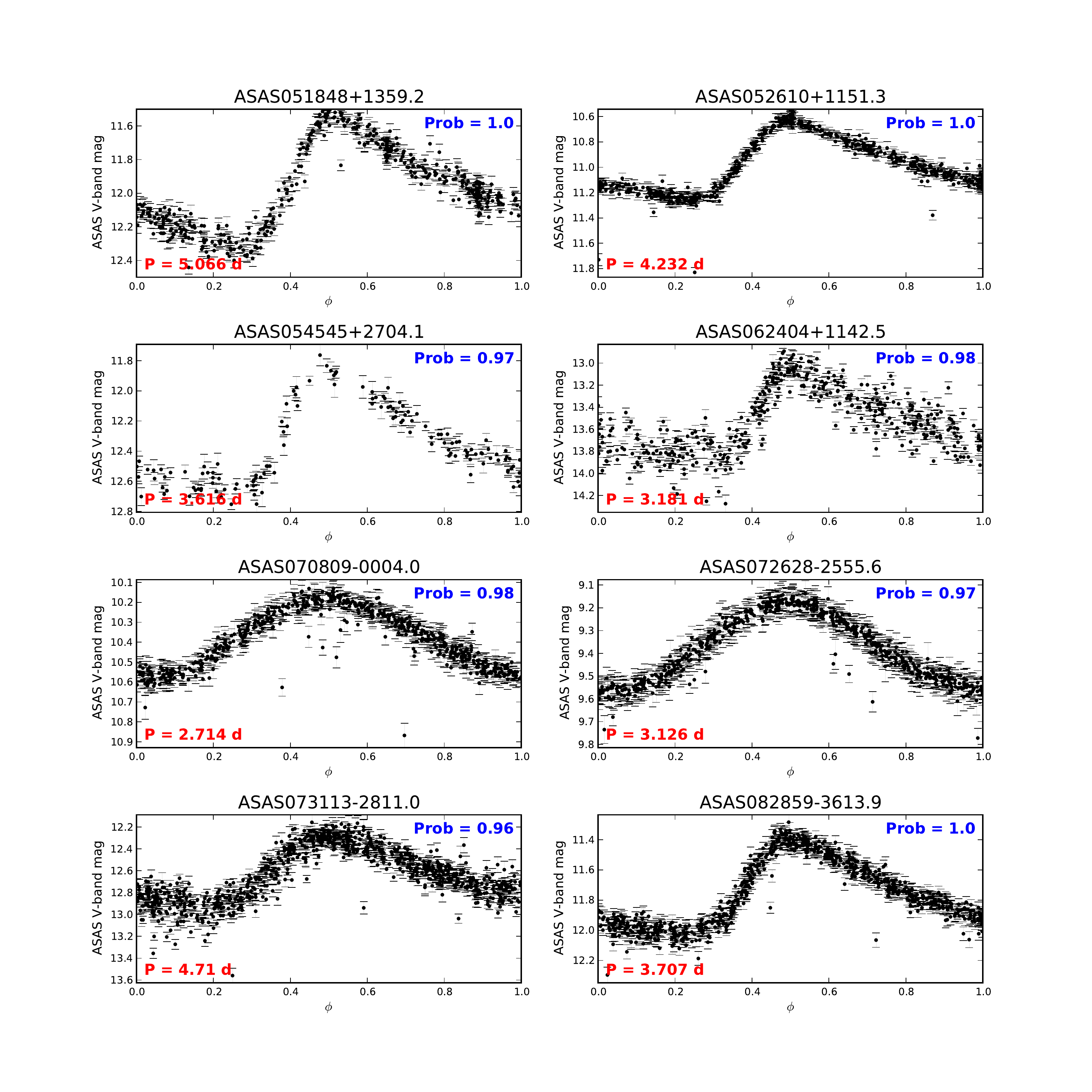} \\
\multicolumn{1}{l}{\mbox{\bf (c)}} & \multicolumn{1}{l}{\mbox{\bf (d)}} \\ 	
\includegraphics[angle=0,width=3.5in]{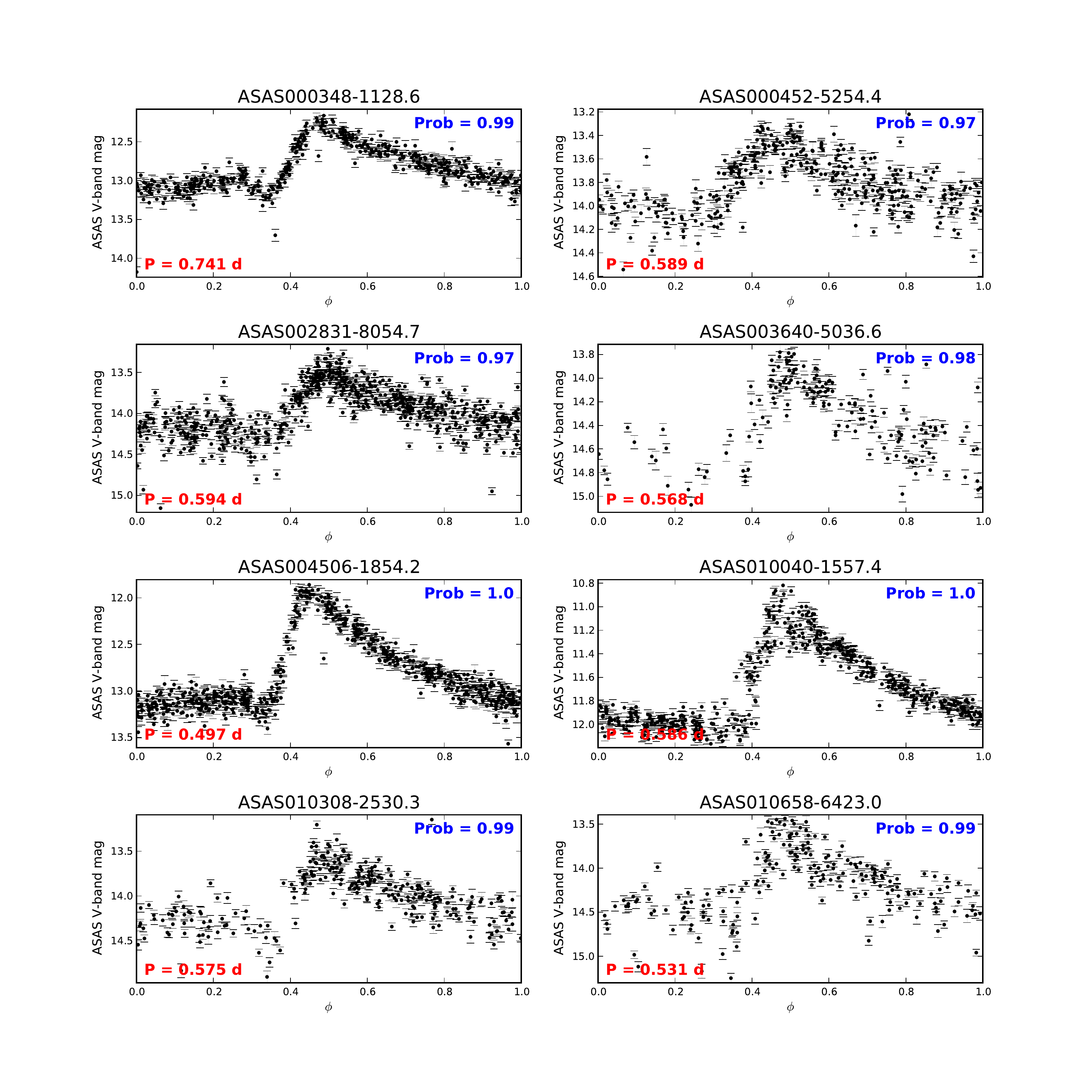} & \includegraphics[angle=0,width=3.5in]{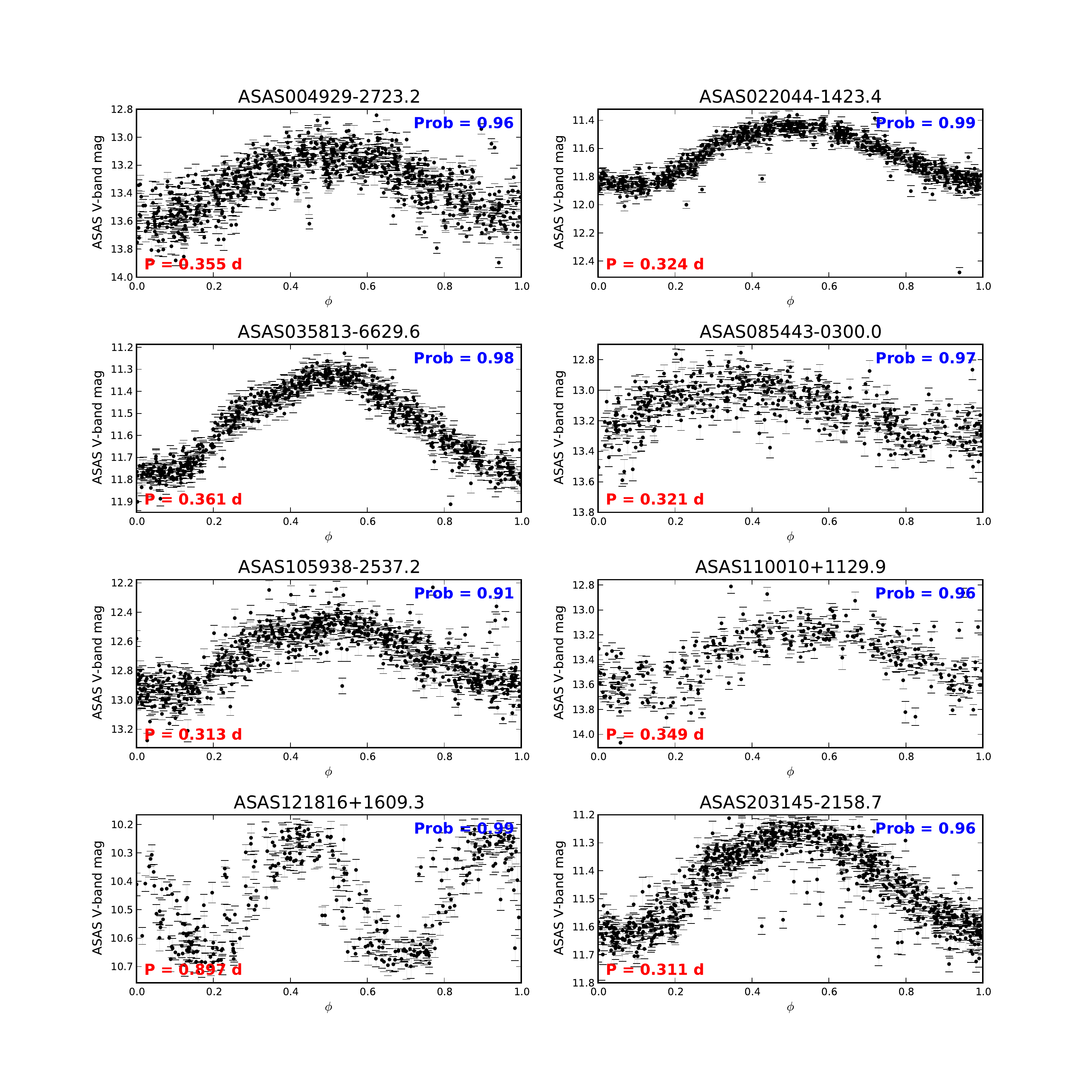} 
\end{array}$
\end{center}
\caption{ ASAS light curves for arbitrarily chosen candidates with high probability of being (a) Mira, (b) Classical Cepheid, (c) Fundamental Mode RR Lyrae, and (d) First Overtone RR Lyrae whose ACVS classification either includes multiple classes, is insecure or MISC, or otherwise differs from that of MACC.  \label{fig:mira} }\end{figure}

\begin{figure}
\begin{center}
$\begin{array}{c@{\hspace{-.3in}}c}
\multicolumn{1}{l}{\mbox{\bf (e)}} & \multicolumn{1}{l}{\mbox{\bf (f)}} \\ 	\vspace{-.1in}
\includegraphics[angle=0,width=3.5in]{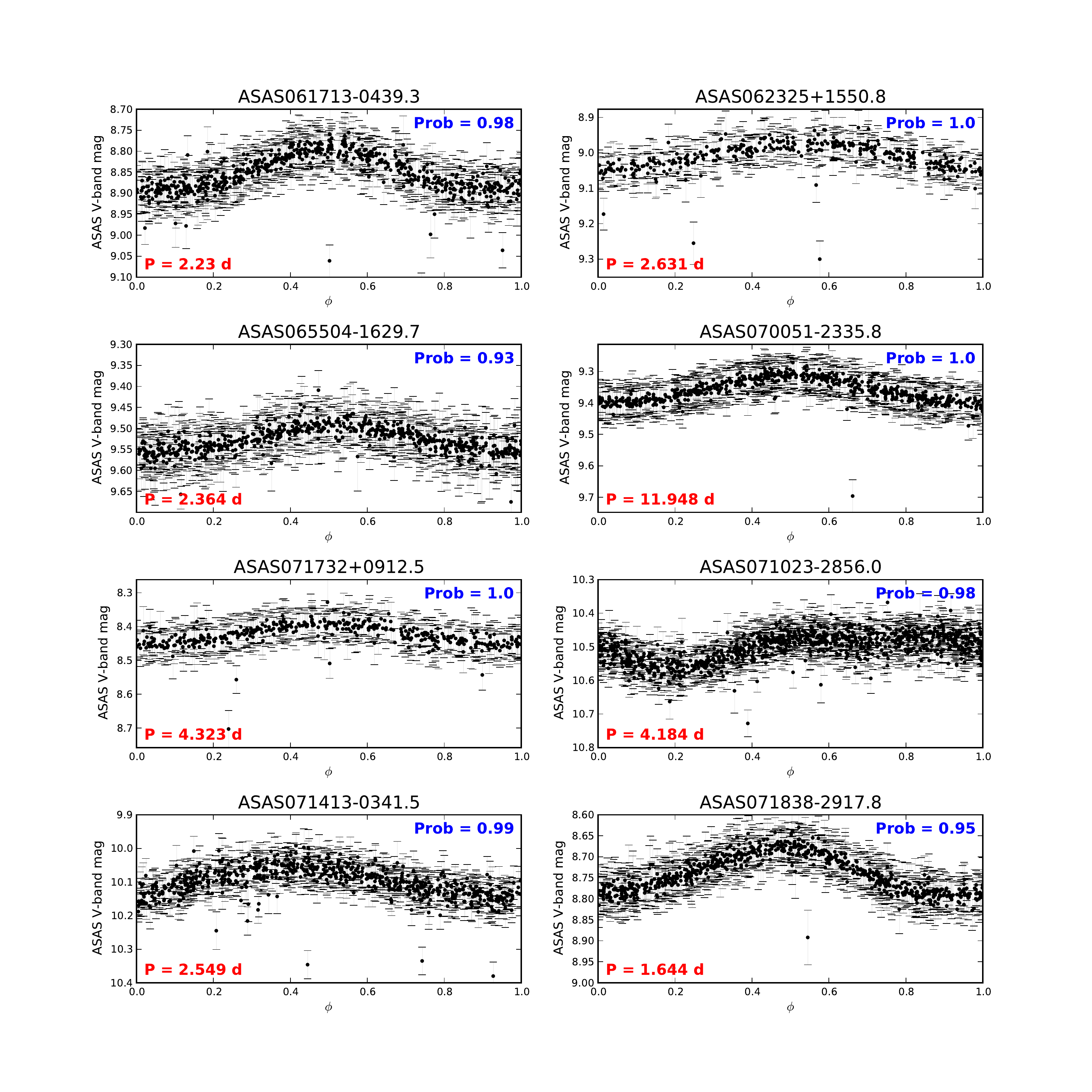} & \includegraphics[angle=0,width=3.5in]{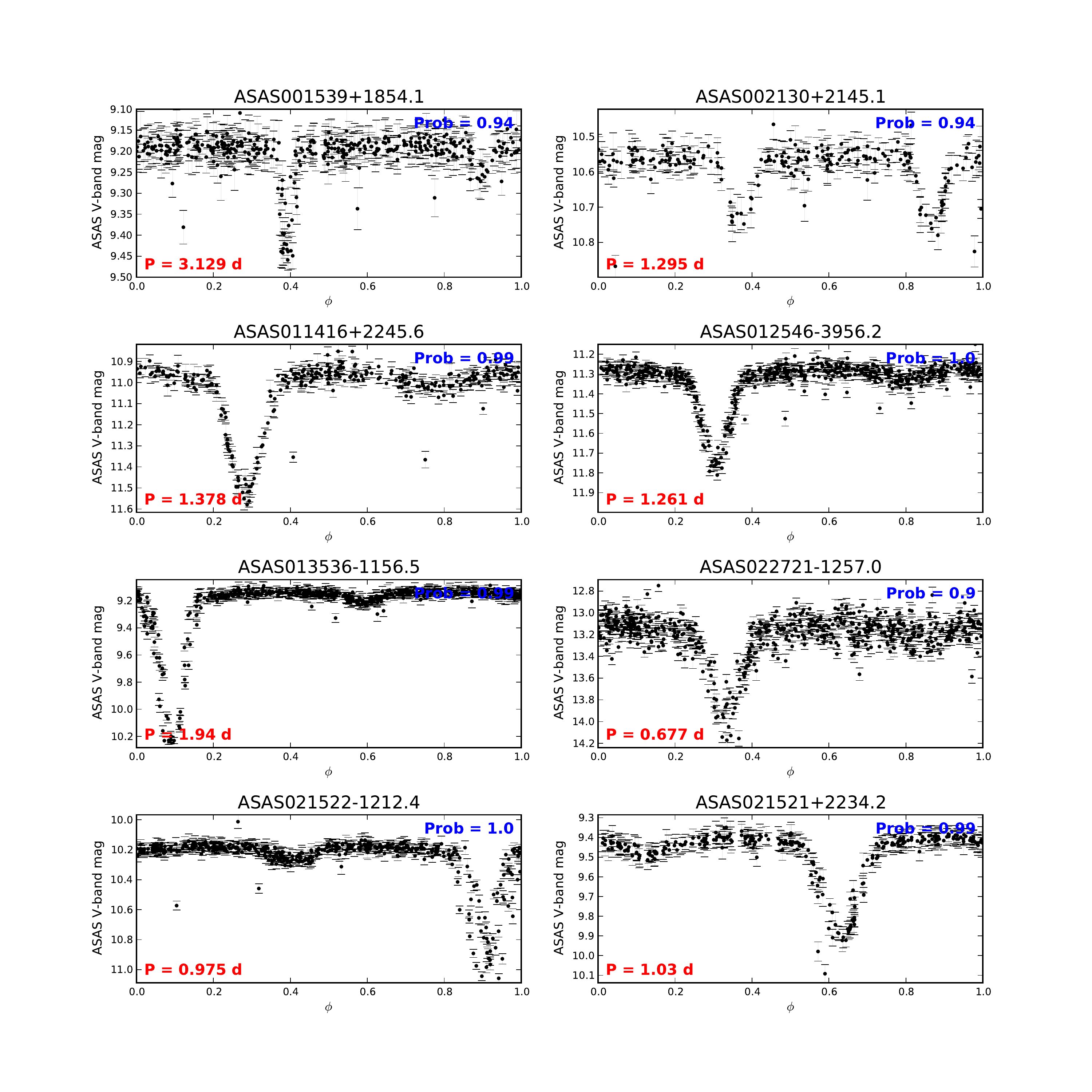} \\
\multicolumn{1}{l}{\mbox{\bf (g)}} & \multicolumn{1}{l}{\mbox{\bf (h)}} \\ 	
\includegraphics[angle=0,width=3.5in]{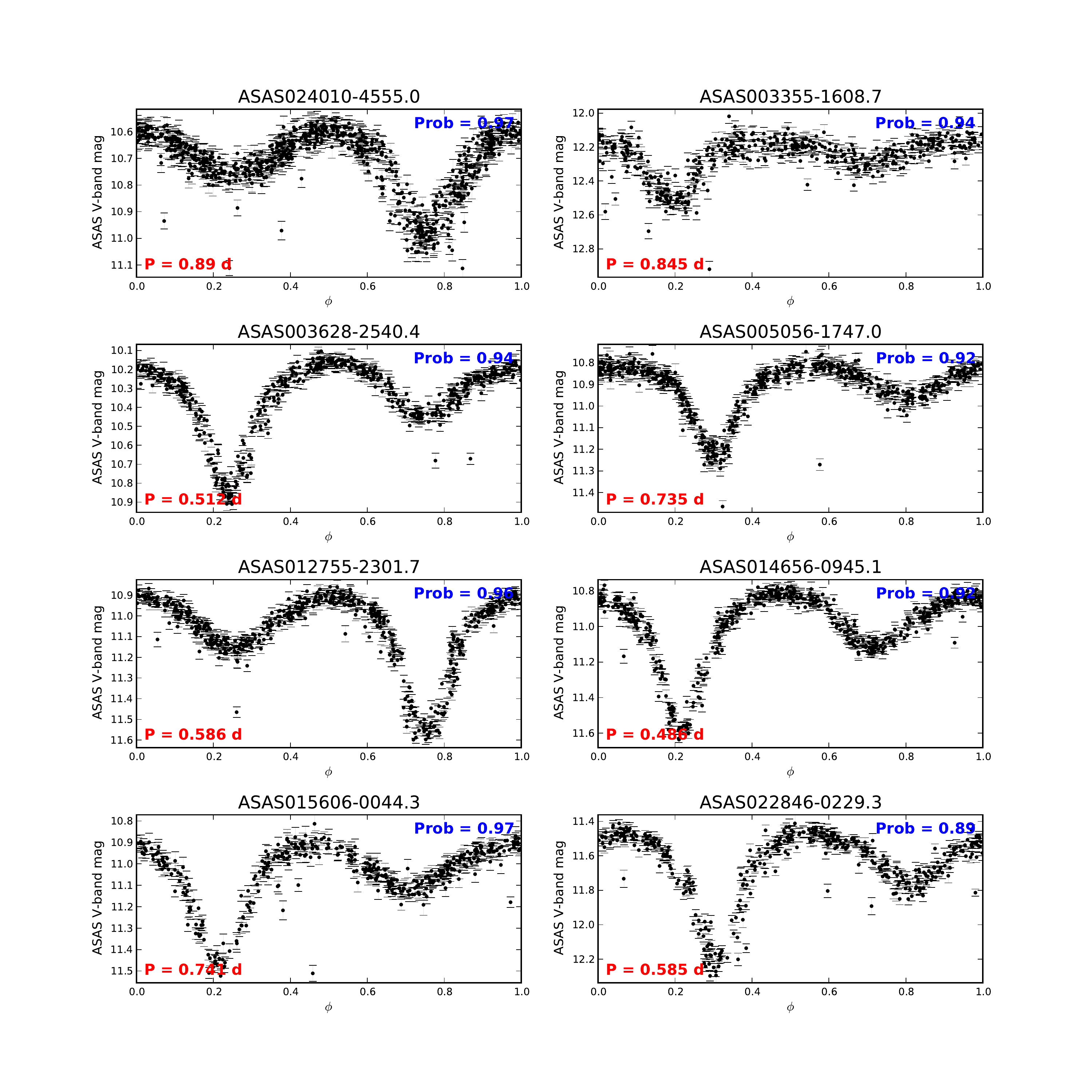} & \includegraphics[angle=0,width=3.5in]{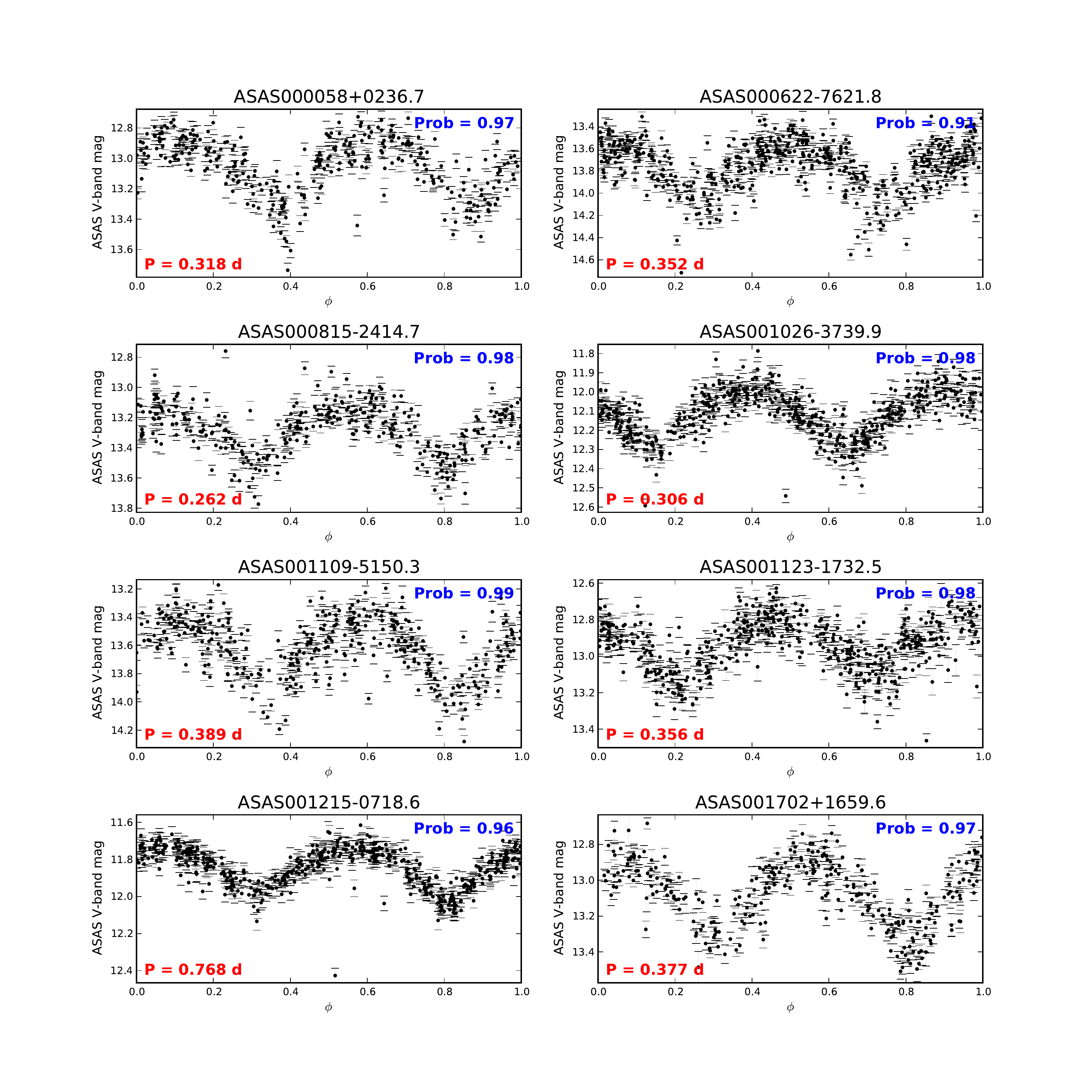}
\end{array}$
\end{center}
\caption{ Same as Figure \ref{fig:mira} for: (e) Chemically Peculiar, (f) Beta Persei, (g) Beta Lyrae, and (h) W Uma Majoris.   \label{fig:wuma} }\end{figure}

In Figures \ref{fig:mira}--\ref{fig:wuma} we plot, for 8 different science classes, the ASAS light curves of selected Machine-learned ASAS Classification Catalog sources whose maximal class probabilities of belonging to that class are greater than 0.9 but whose ACVS classification is different or inconfident.   Within each of these classes, the light curves appear as expected for each class of variability.  MACC is better able to discover the classes of objects near the magnitude limit of ASAS and whose light curves are of lower SNR.

\subsection{Classical Cepheids: \citet{2011ARep...55..816B}}

\citet{2011ARep...55..816B} present multi-band light curves of 49 Classical Cepheid candidates from the ACVS catalog, with data from the 76-cm telescope of the South African Astronomical Observatory and the 40-cm telescope of the Cerro Armazones Observatory of the Catholic University of the North, Chile.  From these observations, they are able to confirm that 48 are Classical Cepheids and one, ASAS 100914-5714.6, is a Double-Mode Cepheid.  Our classifier correctly identifies 46 of these 48 Classical Cepheids.  See Table \ref{tab:cceph} for a complete listing of our catalog classification, posterior probability of Classical Cepheid, ranking of Classical Cepheid probability out of all 50K ASAS sources, and anomaly score for all 49 objects observed by \citet{2011ARep...55..816B}.  None of these objects is in our MACC training set.

For two of these sources, ASAS 075750-2923.5 and ASAS 164120-4739.6, our catalog identifies the objects as Multi-Mode Cepheids.  For the former, which has a period of 2.586 days, there is significant scatter in the phased ASAS light curve and a relatively low amplitude, making its ASAS light curve more consistent with a Multi-Mode Cepheid.  It is likely the presence of a bright neighbor to ASAS 075750-2923.5 that causes this scatter and depressed amplitude.  However, the light curve of \citet{2011ARep...55..816B}, which only contains 9 epochs of data, does not completely rule out a Multi-Mode pulsator.  For the latter, ASAS 164120-4739.6, the object is in the plane (Galactic latitude $-0.842^\circ$) and has a bright neighbor, which again causes a large amount of scatter and suppressed amplitude in the ASAS light curve. Additionally, we find two outliers in the in the \citet{2011ARep...55..816B} catalog, ASAS 073453-2651.3 and ASAS 100914-5714.6, whose Anomaly Scores are greater than the outlier threshold of $t^* = 10.0$ (\S \ref{ss:anomscore}).  The object ASAS 100914-5714.6 is the Double-Mode Cepheid confirmed by \citet{2011ARep...55..816B}, and appears as an outlier due to its anomalously large amplitude and long period for a Double-Mode Cepheid.  The other outlier, ASAS 073453-2651.3, has a median brightness fainter than 14th magnitude, making it barely detectable in ASAS which results in noisy light curve and underestimated amplitude.

\begin{deluxetable}{cccccc} 
\tablecolumns{6} 
\tabletypesize{\tiny}
\tablewidth{0pc} 
\tablecaption{Classification catalog results for Classical Cepheid stars confirmed by \citet{2011ARep...55..816B}.}
\tablehead{ 
\colhead{ASAS ID} & \colhead{Predicted Class} & \colhead{P(Classical Cepheid)} & \colhead{Rank CCeph} & \colhead{Anomaly Score} & \colhead{In Training}}
\startdata 
\input{cceph_result.dat}
\enddata
\label{tab:cceph}
\end{deluxetable}%

\subsection{Beta Cephei: \citet{2005AcA....55..219P}}

In the work of \citet{2005AcA....55..219P}, 14 new Beta Cephei stars appearing in ACVS were confirmed (in addition to 4 other previously known Beta Cephei stars).  Starting with all 37 stars whose ACVS classification includes BCEP as a possible class, the author makes selection cuts based on the ASAS periodogram and any available multi-band photometry and/or spectral type, finding 14 stars that the author deems as unambiguous.  Then, with a broader set of 1700 ASAS stars, Pigulski detects 4 more bona fide candidates using the same selection criteria.

In Table \ref{tab:betaceph} we report our catalog's classification for each of the \citet{2005AcA....55..219P} Beta Cephei.  Note that all but one of these sources was included in the MACC training set.  We mis-identify as a Delta Scuti star the one object (ASAS 161858-5103.5) that was not included in the training set.  This star is located directly in the galactic plane with Galactic latitude of $-0.536^{\circ}$, and suffers from heavy extinction.  Thus its observed colors are more typical of the comparatively redder class of Delta Scuti stars than the bluer class of Beta Cephei.  With a Beta Cephei posterior class probability of 0.243, it ranks within the top 400 Beta Cephei candidates.

\begin{deluxetable}{cccccc} 
\tablecolumns{6} 
\tabletypesize{\footnotesize}
\tablewidth{0pc} 
\tablecaption{Classification catalog results for Beta Cephei stars in \citet{2005AcA....55..219P}.} 
\tablehead{ 
\colhead{ASAS ID} & \colhead{Predicted Class} & \colhead{P(Beta Cephei)}  & \colhead{Rank BetCep} & \colhead{Anomaly Score} & \colhead{In Training} }
\startdata 
\input{betaceph_result.dat}
\enddata
\label{tab:betaceph}
\end{deluxetable}%

\subsection{Double-Mode RR Lyrae: \citet{2007MNRAS.377.1263S}}

\citet{2007MNRAS.377.1263S} perform a search for multiple-pulsating RR Lyrae stars in ASAS.  Starting with all objects with a RR Lyrae classification in ACVS, this study first culled out obvious non-RR Lyrae stars via visual inspection.  They pre-whiten each ASAS RR Lyrae light curve at the pulsation period and run the {\tt CLEAN} algorithm to find any significant periodicity in the residual light curves.  From this analysis, they identify of order 150 Blazhko affected RR Lyrae and 19 Double-Mode RR Lyrae stars.  The Double-Mode pulsators were identified by making cuts on the pulsation period ($P_0$) and the ratio of the overtone to fundamental periods ($ 0.735 \le P_1 / P_0 \le 0.755$) and confirmed via visual inspection.

The MACC classification, posterior probability of Double-Mode RR Lyrae, ranking of RRd amongst all ASAS sources, and anomaly score  for the 19 confirmed RRd from \citet{2007MNRAS.377.1263S} are in Table \ref{tab:rrd}.  MACC correctly classifies all 19 stars even though only two of them were in our training set.  Each of the stars has posterior probability of being a Double-Mode RR Lyrae of $> 0.45$ and each ranks within the top 41 RRd candidates.

\begin{deluxetable}{cccccc} 
\tablecolumns{6} 
\tabletypesize{\footnotesize}
\tablewidth{0pc} 
\tablecaption{Classification catalog results for double-mode RR Lyrae stars in \citet{2007MNRAS.377.1263S}.} 
\tablehead{ 
\colhead{ASAS ID} & \colhead{Predicted Class} & \colhead{P(RRd)} & \colhead{Rank RRd} & \colhead{Anomaly Score} & \colhead{In Training} }
\startdata 
\input{rrldm_result.dat}
\enddata
\label{tab:rrd}
\end{deluxetable}%

\subsection{Orion Belt Objects: \citet{2010AN....331..257C}}

In a search for high-amplitude variable stars in the Orion Belt, \citet{2010AN....331..257C} identify 32 variable stars from ASAS photometry, proper motions, and infrared photometry (2MASS and the Infrared Astronomical Satellite ({\it IRAS})).  They perform an extensive literature search on these objects and visual analysis to determine a likely classification for each.  Of these 32 variable stars, 13 are in our catalog, and their classifications are listed in Table \ref{tab:orion}.  Our classifications agree with those of \citet{2010AN....331..257C} for 9 of the 13 objects.  

For four objects, we disagree with the classifications of \citet{2010AN....331..257C}.   The star ASAS 053621-0210.9 (PQ Ori) was found by us to be a semi-detached (Beta Lyrae) eclipsing system, while \citet{2010AN....331..257C} note that although it has been identified as a possible young stellar object in the literature,  its colors are too blue and it is more likely a field star.  The star ASAS 053946-0055.9 was identified by us as either a LSP or RS CVn, consistent with the classification of \citet{2009OEJV..105....1S}, while \citet{2010AN....331..257C} retain it as an uncertain T Tauri candidate.  The star ASAS 053543-0034.6 is claimed by \citet{2010AN....331..257C} to have signs of youth; however, we find significant periodicity on 86.61-day time scales, which is consistent with the pulsations of a RV Tauri star.  Finally, ASAS 053642+0038.5 is identified by our catalog as a likely W Ursae Majoris candidate due to it's tell-tale eclipsing structure on 1.06-day time scales; \citet{2010AN....331..257C} claim that it is a possible HAeBe star, though they note that it has anomalous brightness.

\begin{deluxetable}{cccccl} 
\tablecolumns{6} 
\tabletypesize{\footnotesize}
\tablewidth{0pc} 
\tablecaption{Classification catalog results for Orion Belt variables in \citet{2010AN....331..257C}.} 
\tablehead{ 
\colhead{ASAS ID} & \colhead{Predicted Class} & \colhead{P(Class)} & \colhead{Anomaly Score} & \colhead{In Training} & \colhead{Caballero Class}}
\startdata 
\input{orion_result.dat}
\enddata
\label{tab:orion}
\end{deluxetable}%

\section{Conclusions}
\label{sec:conclusions}

We have presented an end-to-end methodology for creating a probabilistic classification catalog for a time-domain survey of variability.  With growing data volumes and rates, these types of automated classification catalogs become necessary for astronomers to make sense of such a vast amount of data and to optimize the allocation of limited follow-up resources.  Though machine-learned construction of accurate classification catalogs is certainly a difficult undertaking, we have shown that sub-20\% error rates are achievable even with as many as 28 classes and sub-classes of stellar variability.  Furthermore, we have motivated the importance of disseminating probabilistic classifications with full disclosure of class priors, allowing each user freedom to trade class purity for efficiency and to use full probability vectors in performing astrophysical inference (for a recent use of probabilities for cosmological parameter estimation, see \citealt{2012MNRAS.421..913N}).  Additionally, it is crucial that the classification probabilities be calibrated so that the natural interpretation of probability holds, allowing for faithful propagation of that information to downstream analyses.

As a test case for the methodologies presented in this paper, and
those adopted from \citet{2011rich} and \citet{2012ApJ...744..192R},
we build and make publicly available a 28-class Machine-learned ASAS
Classification Catalog of 50,124 sources that are included in the ASAS
Catalog of Variable Stars.  We show that accurate classifications are
possible for such a complex, noisy and diverse data set of photometric
light curves.  Furthermore, we demonstrate that calibrated
probabilities are attainable using straightforward methodology and
that semi-supervised anomaly detection can discover interesting
objects that do not fit within a predefined classification taxonomy.
Comparisons of our MACC with existing ASAS classifications, including
those in ACVS, are favorable and we eagerly await more intense
scrutiny of the publicly available MACC from the astronomical
community. Inevitably many of our top classifications will be proven
incorrect, but that is expected by the very nature of the product: it
is, instead, the testing of the aggregate accuracy of our
probabilistic classifications that are of most interest long term.

Some degrees of the predicted accuracy and functionality of the MACC catalog have already been demonstrated in the concurrently submitted paper of \citet{rcb}.  In that paper,  MACC was used to search for previously unknown R Coronae Borealis and DY Persei stars in ASAS.  Their search through the top MACC RCB candidates yielded 12 likely RCB/DYPers stars, whereby they confirmed with  new and archival spectroscopic observations the discovery of four RCB stars and four DYPers, increasing the number of known Galactic DYPers from 2 to 6.  \citet{rcb} demonstrate that the MACC catalog recovers ASAS candidates that would have been missed via the typical search method which uses hard cuts on the amplitude and periodicity of the light curves, and that a prohibitive number of objects would have to be manually searched via those traditional methods to recover all of the newly discovered objects.  This is powerful validation that ML probabilistic classification can facilitate astronomical discovery and enable scientific results.

Moving forward, there remain many pending tasks for our machine-learned approach to classification catalogs.  First, we have not touched on the question of discovery of variability, only on classification once variable objects have been identified.  Recently, \citet{2009MNRAS.400.1897S} have introduced a machine learning approach to variability selection which we will expand to develop new procedures.  Second, the size and scope of MACC, at 50k variable stars at a brightness level reaching 14th magnitude, is rather small and limited.  Tackling larger catalogs with millions of sources will test the feasibility of our algorithms and robustness of our statistical approaches.  Third, the future of time-domain surveys is multi-band light curves (e.g., DES, LSST).  Neglecting the full use and exploitation of multi-band photometry would mean throwing away much useful information.  Last, a large component of the catalog-building techniques that we have presented is the constant feedback from the automated classifier and the astronomical community.  From compiling large and representative training sets to inventing new features that probe different types of variability, constant injection of more information into the machine learner is essential to optimize the accuracy, information gain, and ultimately the scientific impact of the catalog.

\acknowledgements

 The authors acknowledge the generous support of a CDI grant (\#0941742) from the National Science Foundation.  This work was performed in the CDI-sponsored Center for Time Domain Informatics (\url{http://cftd.info}).  We would also like to thank IBM and \href{http://citris-uc.org/}{CITRIS} for providing the 280 core cluster at Berkeley, which was used to perform feature computations and classifier evaluations.

\bibliography{TSclassify}

\end{document}

%% file: training_set.dat
Mira & 164 & 0.0843\\
Semireg PV & 101 & 0.0519\\
SARG A & 15 & 0.0077\\
SARG B & 29 & 0.0149\\
LSP & 54 & 0.0278\\
RV Tauri & 25 & 0.0129\\
Classical Cepheid & 204 & 0.1049\\
PopII Cepheid & 27 & 0.0139\\
MultiMode Cepheid & 98 & 0.0504\\
RR Lyrae FM & 148 & 0.0761\\
RR Lyrae FO & 39 & 0.0201\\
RR Lyrae DM & 59 & 0.0303\\
Delta Scuti & 133 & 0.0684\\
SX Phe & 6 & 0.0031\\
Beta Cephei & 55 & 0.0283\\
Pulsating Be & 49 & 0.0252\\
PerVarSG & 55 & 0.0283\\
ChemPeculiar & 75 & 0.0386\\
RCB & 17 & 0.0087\\
ClassT Tauri & 12 & 0.0062\\
Weakline T Tauri & 20 & 0.0103\\
RS CVn & 17 & 0.0087\\
Herbig AEBE & 22 & 0.0113\\
S Doradus & 7 & 0.0036\\
Ellipsoidal & 13 & 0.0067\\
Beta Persei & 178 & 0.0915\\
Beta Lyrae & 202 & 0.1039\\
W Ursae Maj & 121 & 0.0622\\

%% file: asas_catalog.dat
000006+2553.2 & 215153 & 0.027375 & 25.886453 & Mira & 1 & 0.035 & MIRA & Mira & 1 &  ...   & 319.295 & 10.398 & 170 & 8.56 & 2.488\\
000007+1844.3 & 215154 & 0.030375 & 18.738077 & Beta Persei & 0.834 & 1.646 & ESD/CW$$-$$FU/ACV/ED &  & 0.001 &  ...  & 2.589 & 10.795 & 304 & 10.85 & 0.41\\
000007+2014.3 & 215155 & 0.028755 & 20.237385 & Semireg PV & 0.976 & 1.404 & MISC &  & 0.002 &  ...  & 213.732 & 9.064 & 233 & 9.07 & 0.558\\
000017+2636.4 & 215156 & 0.068685 & 26.608939 & Mira & 0.382 & 5.579 & MIRA &  & 0.382 &  ...   & 186.65 & 8.126 & 119 & 10.83 & 1.149\\
000018+0919.4 & 215157 & 0.075405 & 9.323315 & RS CVn & 0.287 & 7.197 & MISC &  & 0.004 &  ...   & 43.541 & 10.755 & 239 & 10.52 & 0.108\\
000030$-$3937.5 & 215158 & 0.129525 & $-$39.630347 & Beta Persei & 0.652 & 2.571 & ED &  & 0.001 &  ...  & 2.553 & 9.641 & 410 & 10.79 & 0.614\\
000036+2639.8 & 215159 & 0.14772 & 26.663685 & RR Lyrae FM & 0.993 & 0.534 & RRAB &  & 0 &  ...   & 0.566 & 6.65 & 87 & 12.61 & 0.578\\
000053$-$1717.5 & 215160 & 0.220095 & $-$17.291745 & W Ursae Maj & 0.905 & 1.618 & ESD/EC &  & 0 &  ...   & 0.298 & 13.769 & 345 & 12.66 & 0.524\\
000058+0236.7 & 215161 & 0.23553 & 2.611892 & W Ursae Maj & 0.972 & 1.439 & EC/DSCT/ESD &  & 0 &  ...   & 0.318 & 11.915 & 296 & 12.94 & 0.504\\
000058+1236.5 & 215162 & 0.242265 & 12.607833 & Weakline T Tauri & 0.264 & 11.821 & MISC &  & 0.013 &  ...   & 5.204 & 7.917 & 237 & 12.64 & 0.348\\
000108$-$3330.1 & 215163 & 0.28191 & $-$33.500881 & W Ursae Maj & 0.976 & 0.779 & EC &  & 0 &  ...   & 0.467 & 16.315 & 425 & 11.51 & 0.336\\
000112+0904.7 & 215164 & 0.297765 & 9.078171 & Delta Scuti & 0.608 & 2.289 & ESD &  & 0 &  ...   & 0.121 & 11.002 & 240 & 10.28 & 0.08\\
000116$-$6037.0 & 215165 & 0.316515 & $-$60.615788 & Delta Scuti & 0.847 & 3.132 & DSCT &  & 0 &  ...  & 0.122 & 15.126 & 536 & 10.03 & 0.356\\
000118$-$3551.7 & 215166 & 0.32487 & $-$35.860717 & SARG A & 0.547 & 2.333 & MISC &  & 0 &  ...   & 25.488 & 8.803 & 451 & 9.84 & 0.254\\
000119$-$3505.9 & 215167 & 0.32922 & $-$35.097789 & SARG B & 0.91 & 1.513 & MISC &  & 0 &  ...   & 38.742 & 7.956 & 437 & 10.77 & 0.258\\
000120$-$5834.8 & 215168 & 0.334545 & $-$58.580264 & LSP & 0.677 & 2.759 & MISC &  & 0.006 &  ...   & 375.134 & 11.847 & 475 & 9.53 & 0.318\\
000139$-$0345.4 & 215169 & 0.418155 & $-$3.756766 & Semireg PV & 0.965 & 1.604 & MISC &  & 0.003 &  ...   & 162.98 & 11.858 & 344 & 12.82 & 0.898\\
000142$-$4229.3 & 215170 & 0.425685 & $-$42.487419 & SARG B & 0.475 & 4.495 & MISC &  & 0.008 &  ...   & 96.313 & 8.014 & 424 & 10.71 & 0.379\\
000147$-$5714.5 & 215171 & 0.44919 & $-$57.242031 & Delta Scuti & 0.594 & 3.425 & ESD/EC &  & 0 &  ...   & 0.235 & 17.212 & 475 & 11.03 & 0.15\\
000155$-$6707.7 & 215172 & 0.475245 & $-$67.130487 & RV Tauri & 0.361 & 7.333 & MISC &  & 0.085 &  ...   & 1.006 & 6.464 & 217 & 12.74 & 1.335\\
000157$-$5250.1 & 215173 & 0.48729 & $-$52.835239 & LSP & 0.534 & 3.167 & MISC &  & 0.021 &  ...   & 2109.092 & 11.879 & 532 & 10.78 & 0.419\\
000158+1357.6 & 215174 & 0.49407 & 13.959879 & Weakline T Tauri & 0.376 & 5.024 & MISC &  & 0.005 &  ...   & 74.32 & 7.228 & 191 & 11.52 & 0.16\\
000202$-$6653.3 & 215175 & 0.498525 & $-$66.882596 & W Ursae Maj & 0.992 & 0.901 & EC &  & 0 &  ...   & 0.327 & 16.748 & 462 & 12.16 & 0.501\\
000208$-$1440.5 & 215176 & 0.532755 & $-$14.674641 & Mira & 0.999 & 0.163 & MIRA &  & 0.999 &  ...   & 351.44 & 15.273 & 358 & 8.35 & 3.534\\
000221$-$2929.6 & 215177 & 0.582285 & $-$29.493562 & Weakline T Tauri & 0.333 & 10.628 & ESD/ED &  & 0.005 &  ...   & 3.143 & 13.906 & 448 & 12.32 & 0.462\\
000222+0429.6 & 215178 & 0.58593 & 4.494017 & PerVarSG & 0.306 & 7.772 & MISC &  & 0.001 &  ...  &  307.745 & 4.87 & 302 & 9.79 & 0.134\\
000229$-$5653.9 & 215179 & 0.61572 & $-$56.898676 & ChemPeculiar & 0.294 & 6.463 & ESD/EC/ELL/SR &  & 0 &  ...   & 13.223 & 13.931 & 474 & 10.15 & 0.124\\
000239$-$1926.7 & 215180 & 0.668595 & $-$19.442961 & SARG B & 0.513 & 4.051 & MISC &  & 0.004 &  ...   & 453.892 & 7.716 & 442 & 9.69 & 0.346\\
000248$-$2456.7 & 215181 & 0.70044 & $-$24.945325 & RR Lyrae FM & 1 & 0.121 & RRAB & RR Lyrae FM & 0 &  ...  & 0.493 & 16.031 & 387 & 9.9 & 0.704\\
000301$-$7041.5 & 215182 & 0.751575 & $-$70.685852 & RR Lyrae FM & 0.997 & 0.33 & RRAB &  & 0 &  ...  & 0.554 & 13.126 & 340 & 13.27 & 0.838\\
000309$-$1050.5 & 215183 & 0.78684 & $-$10.841496 & Semireg PV & 0.596 & 4.556 & MISC &  & 0.014 &  ...  & 75.275 & 8.376 & 329 & 12.01 & 0.548\\
000309$-$3456.3 & 215184 & 0.78795 & $-$34.935204 & W Ursae Maj & 0.998 & 0.497 & EC &  & 0 &  ...   & 0.265 & 15.474 & 428 & 12.4 & 0.482\\
000316$-$1125.1 & 215185 & 0.817815 & $-$11.417582 & RR Lyrae FO & 0.349 & 5.41 & RRAB: &  & 0 &  ...   & 0.369 & 6.269 & 225 & 13.64 & 0.654\\
000316$-$7342.2 & 215186 & 0.817185 & $-$73.701505 & W Ursae Maj & 0.572 & 2.759 & DSCT/EC &  & 0 &  ...   & 0.201 & 11.158 & 345 & 13.72 & 0.608\\
000321+0323.9 & 215187 & 0.83778 & 3.398293 & RR Lyrae FM & 0.996 & 0.548 & RRAB &  & 0 &  ...  & 0.579 & 10.058 & 208 & 13.21 & 0.612\\
000336$-$1452.4 & 215188 & 0.901605 & $-$14.873255 & W Ursae Maj & 0.61 & 5.173 & EC &  & 0 &  ...  & 1.055 & 14.752 & 358 & 10.93 & 0.268\\
000341$-$3906.9 & 215189 & 0.918975 & $-$39.113062 & Semireg PV & 0.539 & 7.197 & MISC &  & 0.01 &  ...  & 0.998 & 9.871 & 424 & 11.65 & 0.608\\
000348$-$1128.6 & 215190 & 0.952425 & $-$11.475164 & RR Lyrae FM & 0.993 & 0.404 & RRAB/DCEP$-$FO &  & 0 &  ...  & 0.741 & 15.445 & 401 & 12.45 & 0.77\\
000358+0223.1 & 215191 & 0.99306 & 2.385359 & SARG A & 0.307 & 6.937 & MISC &  & 0 &  ...  & 18.085 & 6.467 & 318 & 9.9 & 0.114\\
000405$-$1659.8 & 215192 & 1.021365 & $-$16.997633 & RR Lyrae FM & 0.999 & 0.33 & RRAB & RR Lyrae FM & 0 &  ...  & 0.606 & 14.636 & 350 & 11.94 & 0.588\\

%% file: cceph_result.dat
052610+1151.3 & Classical Cepheid & 0.9995 & 45 & 0.185 & no\\
052706+1656.2 & Classical Cepheid & 0.9591 & 179 & 1.165 & no\\
062939-1840.5 & Classical Cepheid & 0.9544 & 185 & 1.242 & no\\
064037+1143.6 & Classical Cepheid & 0.98 & 142 & 0.976 & no\\
064829-1014.2 & Classical Cepheid & 0.9246 & 217 & 1.488 & no\\
070355-1752.8 & Classical Cepheid & 0.988 & 126 & 0.859 & no\\
071342-1737.2 & Classical Cepheid & 0.9284 & 214 & 1.89 & no\\
073113-2811.0 & Classical Cepheid & 0.9639 & 170 & 1.525 & no\\
073453-2651.3 & Classical Cepheid & 0.2799 & 559 & 12.158 & no\\
073502-3554.9 & Classical Cepheid & 0.915 & 226 & 1.66 & no\\
074925-3814.4 & Classical Cepheid & 0.9733 & 155 & 0.866 & no\\
075345-3658.2 & Classical Cepheid & 0.9997 & 39 & 0.235 & no\\
075358-2822.1 & Classical Cepheid & 0.8161 & 277 & 1.525 & no\\
075750-2923.5 & MultiMode Cepheid & 0.0899 & 789 & 3.808 & no\\
075840-3330.2 & Classical Cepheid & 0.9927 & 109 & 0.582 & no\\
075912-2641.9 & Classical Cepheid & 0.6498 & 341 & 4.102 & no\\
080500-2851.8 & Classical Cepheid & 0.99 & 118 & 0.475 & no\\
080511-3421.7 & Classical Cepheid & 0.9806 & 140 & 0.73 & no\\
080927-3315.7 & Classical Cepheid & 0.9674 & 164 & 1.101 & no\\
081025-3828.4 & Classical Cepheid & 0.9107 & 227 & 1.825 & no\\
081026-3231.3 & Classical Cepheid & 0.9796 & 143 & 0.887 & no\\
082117-3845.3 & Classical Cepheid & 0.9419 & 203 & 0.961 & no\\
082127-3825.3 & Classical Cepheid & 0.9608 & 176 & 1.252 & no\\
082859-3613.9 & Classical Cepheid & 0.997 & 85 & 0.348 & no\\
083130-4429.3 & Classical Cepheid & 0.5067 & 419 & 5.173 & no\\
083426-3559.1 & Classical Cepheid & 0.9936 & 106 & 0.475 & no\\
083611-3903.7 & Classical Cepheid & 0.7172 & 309 & 1.924 & no\\
084127-4353.6 & Classical Cepheid & 0.8537 & 258 & 2.521 & no\\
090436-4633.2 & Classical Cepheid & 0.6235 & 356 & 7.621 & no\\
090932-5359.3 & Classical Cepheid & 0.9688 & 162 & 1.024 & no\\
092758-5218.9 & Classical Cepheid & 0.9592 & 178 & 1.083 & no\\
093005-5137.5 & Classical Cepheid & 0.9373 & 209 & 0.961 & no\\
094819-5748.6 & Classical Cepheid & 0.6993 & 316 & 3.167 & no\\
094827-5801.1 & Classical Cepheid & 0.9955 & 96 & 0.453 & no\\
100914-5714.6 & Classical Cepheid & 0.5148 & 415 & 10.905 & no\\
101037-5817.8 & Classical Cepheid & 0.797 & 285 & 2.521 & no\\
101538-5933.1 & Classical Cepheid & 0.4654 & 442 & 4.952 & no\\
103627-6211.6 & Classical Cepheid & 0.8333 & 268 & 1.488 & no\\
112039-6149.9 & Classical Cepheid & 0.8839 & 239 & 2.226 & no\\
115701-6218.7 & Classical Cepheid & 0.3507 & 517 & 9.638 & no\\
122240-6209.5 & Classical Cepheid & 0.858 & 254 & 1.874 & no\\
123804-3831.4 & Classical Cepheid & 0.9529 & 189 & 1.008 & no\\
140742-6315.4 & Classical Cepheid & 0.9268 & 215 & 1.66 & no\\
150547-5823.0 & Classical Cepheid & 0.9529 & 190 & 1.193 & no\\
152021-5807.3 & Classical Cepheid & 0.8242 & 272 & 3.63 & no\\
164120-4739.6 & MultiMode Cepheid & 0.2179 & 576 & 3.95 & no\\
173253-3554.7 & Classical Cepheid & 0.7444 & 297 & 4.435 & no\\
174134-4854.6 & Classical Cepheid & 0.5434 & 398 & 4.814 & no\\
181416-0920.4 & Classical Cepheid & 0.5961 & 369 & 4.263 & no\\

%% file: betaceph_result.dat
091731-5250.3 & Beta Cephei & 0.9767 & 6 & 2.497 & yes\\ 
180233-4005.2 & Beta Cephei & 0.9942 & 1 & 0.923 & yes\\ 
191715+0103.6 & Beta Cephei & 0.9563 & 13 & 3.237 & yes\\ 
212329+0955.9 & Beta Cephei & 0.9547 & 14 & 5.329 & yes\\ 
122213-6320.8 & Beta Cephei & 0.9781 & 5 & 2.03 & yes\\ 
150955-6530.4 & Beta Cephei & 0.9864 & 3 & 2.106 & yes\\ 
161858-5103.5 & Delta Scuti & 0.2428 & 393 & 3.237 & no\\ 
164409-4719.1 & Beta Cephei & 0.9301 & 19 & 2.846 & yes\\ 
164630-4701.2 & Beta Cephei & 0.9436 & 17 & 1.793 & yes\\ 
164939-4431.7 & Beta Cephei & 0.9384 & 18 & 2.356 & yes\\ 
165314-4345.0 & Beta Cephei & 0.9695 & 9 & 2.356 & yes\\ 
165554-4808.8 & Beta Cephei & 0.9703 & 8 & 1.841 & yes\\ 
171218-3306.1 & Beta Cephei & 0.9675 & 11 & 2.356 & yes\\ 
180808-3434.5 & Beta Cephei & 0.9687 & 10 & 2.497 & yes\\ 
181716-1527.1 & Beta Cephei & 0.9807 & 4 & 2.497 & yes\\ 
182610-1704.3 & Beta Cephei & 0.9729 & 7 & 1.551 & yes\\ 
182617-1515.7 & Beta Cephei & 0.9541 & 15 & 5.25 & yes\\ 
182726-1442.1 & Beta Cephei & 0.9526 & 16 & 1.841 & yes\\ 

%% file: rrldm_result.dat
032820-6458.7 & RR Lyrae DM & 0.6938 & 15 & 2.185 & no\\
040054-4923.8 & RR Lyrae DM & 0.8174 & 5 & 1.874 & no\\
081610-6644.8 & RR Lyrae DM & 0.5624 & 29 & 3.425 & no\\
084747-0339.1 & RR Lyrae DM & 0.5394 & 32 & 4.814 & no\\
122509-2139.9 & RR Lyrae DM & 0.4625 & 41 & 4 & no\\
133439+2416.6 & RR Lyrae DM & 0.6745 & 19 & 3.63 & no\\
141539+0010.1 & RR Lyrae DM & 0.7266 & 12 & 3.132 & no\\
151735-0105.3 & RR Lyrae DM & 0.6488 & 22 & 3.31 & no\\
173726+1122.4 & RR Lyrae DM & 0.7197 & 13 & 2.571 & no\\
183952-3200.9 & RR Lyrae DM & 0.9174 & 3 & 6.463 & yes\\
184035-5350.7 & RR Lyrae DM & 0.9682 & 2 & 0.953 & no\\
193933-6528.9 & RR Lyrae DM & 0.8154 & 6 & 2.145 & no\\
195612-5043.7 & RR Lyrae DM & 0.9872 & 1 & 1.11 & yes\\
210726+0110.3 & RR Lyrae DM & 0.6898 & 17 & 2.65 & no\\
211848-3430.4 & RR Lyrae DM & 0.7864 & 10 & 2.676 & no\\
212721-1908.0 & RR Lyrae DM & 0.7902 & 9 & 1.841 & no\\
213437-4907.5 & RR Lyrae DM & 0.6397 & 23 & 3.673 & no\\
230449-3345.3 & RR Lyrae DM & 0.7141 & 14 & 3.132 & no\\
235622-5329.4 & RR Lyrae DM & 0.6855 & 18 & 2.571 & no\\

%% file: orion_result.dat
054354-0243.6 & W Ursae Maj & 0.9993 & 0.33 & no & Contactbinary\\
053848-0227.2 & Weakline T Tauri & 0.3377 & 6.463 & no & TTauri\\
053621-0210.9 & Beta Lyrae & 0.8389 & 3.274 & no & HAeBe\\
053739-0146.3 & Mira & 0.999 & 0.163 & no & Giant\\
053757-0140.8 & Semireg PV & 0.8433 & 2.086 & no & Giant\\
053126-0058.6 & W Ursae Maj & 0.7972 & 2.145 & no & Unknown\\
053946-0055.9 & LSP & 0.326 & 10.111 & no & TTauri?\\
052725-0035.2 & SARG B & 0.5592 & 2.289 & no & Giant\\
053543-0034.6 & RV Tauri & 0.478 & 4.495 & no & TTauri\\
052634-0019.5 & SARG B & 0.8339 & 3.032 & no & Giant\\
054612+0032.4 & ClassT Tauri & 0.3584 & 4.882 & no & Unknown\\
053642+0038.5 & W Ursae Maj & 0.47 & 6.353 & no & HAeBe?\\
053348+0055.6 & SARG B & 0.5954 & 3.167 & no & Giant\\